\documentclass[12pt]{iopart} 
\usepackage{amssymb}

\usepackage[usenames, dvipsnames]{color}

\usepackage{graphicx}

\usepackage[colorlinks=true,linkcolor=black, citecolor=blue]{hyperref}

\newcommand\figwidth{102} 
\newcommand\figwidthtwo{34} 

\begin{document}

\title[]{Asynchronous Lock In Thermography of 3D Printed PLA and ABS samples}

\author[label2]{K. H. H. Goh$^{1}$, Q. F. Lim and P. K. Pallathadka$^{2}$}

\address{Institute of Materials Research and Engineering,

2 Fusionopolis Way. Innovis, \#08-03, Singapore 138634}
\eads{\mailto{henrygohkh@hotmail.com}$^{1}$, \mailto{pramoda-kp@imre.a-star.edu.sg}$^{2}$}

\begin{abstract}
Lock-In thermography is a useful Non Destructive Technique (NDT) for enhanced detection of defects in components, as it amplifies the phase contrast where defects exist. This amplification was found to be around 2-3 times compared to constant heating. The current used a Fuse Deposition Modelling (FDM) 3D printer to print samples with known defects, in order to characterise the relative effects of different variables on the Lock-In phase data. Samples were printed using ABS (Acrylonitrile Butadiene Styrene) and PLA (Polylactic Acid) for comparisons, and variables such as print direction, cameras, heating power, Lock-In frequency, as well as thickness, width and depth of defects were explored. It was found that different materials resulted in different baselines, but had similar phase contrast. A novel asynchronous technique was derived to enable Lock-In measurements with 5 different infrared cameras, and similar results were found. Even cheap cameras like the Seek Thermal CompactXR were proven capable of detecting the same defects as other cameras such as the FLIR SC7500. Heating power did not affect phase contrast, except for shallower defects up to 1.0~mm deep, where higher power resulted in better contrast. As expected, deeper defects could only be detected using lower Lock-In frequencies, and there was better phase contrast with wider, thicker and shallower defects. It was shown that defects 4~mm in width could be detected automatically up to a depth of around 1.5~mm, based on the phase signal trends. Sub-sampling of frame data showed that at least 10 frames were required per Lock-In period for minimal deviations in Lock-In phase contrast. Also, it was shown that phase contrast was similar for shallower defects up to 1.5~mm deep, with data from 1 Lock-In period, as long as the first frame was synchronised with the heating cycle.
\end{abstract}

%
%
\noindent{\it Keywords}: Thermography, 3D Printing, Asynchronous, Lock-In Thermography, ABS, PLA

\submitto{\MST}

%
%

\section{Introduction}
\label{sect:intro}

Non-destructive testing (NDT) is a common approach used to assess and identify defects in structures and components. One of the techniques include infrared thermography (IRT), where infrared signals are used to detect defects, without causing damage to structures and components. Examples of applications of infrared thermography include building diagnostics in terms of mechanical and electrical inspections~\cite{Balaras2002EB}, detection of hidden structures and moisture content in ancient buildings~\cite{Grinzato2002JCH}, detection of water ingress in honeycomb structures used in aviation~\cite{Valilov2016PT}, as well as quality control of electronic components such as solar cells~\cite{Bauer2009EDFA}. There are many different techniques available for infrared thermography, and a good review of existing techniques was presented by Usamentiaga~\etal~\cite{Usamentiaga2014SS}.

Existing techniques use different sources of heat generation with infrared cameras to detect hidden features and defects. These methods use either pulsed heat sources, or modulated heat sources. These are coupled with different signal processing techniques. Pulsed thermography uses short heat pulses created by Xenon flash lamps, and a relatively well known signal processing technique is the Thermographic Signal Reconstruction~(TSR) algorithm~\cite{Balageas2015BBE}. For longer pulsed heating using other sources of heat, the Pulse Phase Thermography~(PPT) technique~\cite{Maldague1996JAP} is used to detect defects from cooling curves. An alternative to Pulse Phase Thermography was introduced by Rajic~\cite{Rajic2002CS}, to process the same cooling curves using Principal Component Thermography~(PCT). In terms of modulated thermography, the Four Point Correlation Method~(FPCM)~\cite{Busse1992JAP} as well as Lock-In Thermography~(LIT)~\cite{Kuo1989SPIE} are the existing signal processing techniques used. Comparisons were also drawn between different techniques from previous studies~\cite{Keo2015CPE,Liu2010IPT}.

It is interesting to note that variations in conventional thermography techniques exist, such as the use of induction loops for real time detection of cracks in steel~\cite{Caiza2015AITA} with Lock-In Thermography, as well as the development of real time Lock-In Thermography~\cite{Schlangen2010MR} to obtain processed images while capturing data. 
It can also be observed that extensive work on thermography has been conducted in the past on different materials such as ceramics~\cite{Sakagami2002IPT,Grinzato2002JCH}, metals\cite{Zimnoch2010AMEA,Maldague2002IPT,Sakagami2002IPT,Caiza2015AITA}, aluminium foam~\cite{Duan2013IPT}, as well as composite structures~\cite{Keo2015CPE,Rajic2002CS,Valilov2016PT,Mulaveesala2006APL,Sakagami1994JSME}.       From the existing literature, it can be concluded that many infrared techniques exist, for a wide variety of materials.

 In order to form useful conclusions, the current study focused on using conventional Lock-In Thermography to detect known defects in polymers. These defects were air pockets of known geometries. Variables that were studied include materials, depth of defects, size of defects, thickness of defects, Lock-In frequencies and heating power. In addition, as the samples were 3D printed using Fused Deposition Modelling (FDM), print directions were also considered as variables. The effects of using different cameras were also taken into consideration, to understand the relative effects of wavelength range of signals and specifications of the cameras. 

\section{Materials and Methods}
\label{sect:methods}

\begin{figure} 
\begin{center}
\includegraphics[width=\figwidth mm]{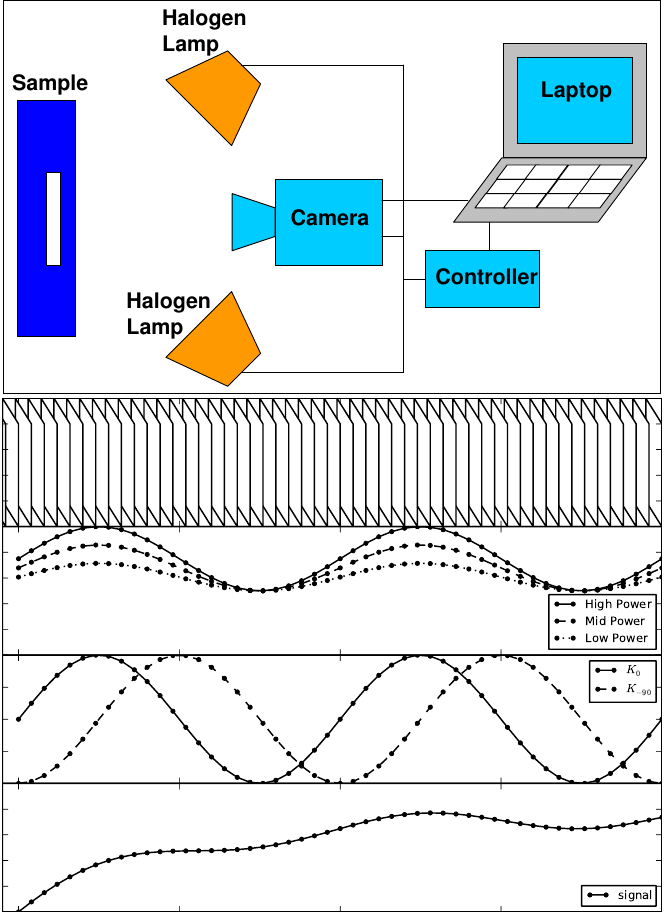}
\end{center}
\caption{Schematic of experimental setup and signal processing techniques. The four plots show the following (top to bottom): (1)~Images stacked into a 3D array. (2)~Relative power from the halogen lamps, from high to low power. (3)~Weighting factors K. (4)~Measured signal profiles from single pixels on the images, plotted over time.
\label{fig:setup}}
\end{figure}

A schematic of the setup is shown in Figure~\ref{fig:setup}. The setup consists of a laptop, different infrared cameras, halogen lamps and a custom built controller. Cameras used in this study include FLIR SC7500, FLIR A310, Fluke Ti55FT, Seek Thermal Compact XR and Xenics Gobi640, with their respective specifications shown in Table~\ref{tab:cameras}. Different synchronisation techniques were used to synchronise the heat source (halogen lamps) to the camera frames. Altair was used for measurements on SC7500, where measurements were synchronised with the heat source using a hardware trigger signal from the controller box. FLIR ResearchIR was used for controlling the A310. The Ti55FT could not be controlled via computers, so images were acquired directly on the camera, and read from the Compact Flash card (CF card) using SmartView software. A customised Labview application was used for controlling the Seek Thermal Compact XR. Xeneth was used for controlling the Xenics Gobi640. Synchronisation was done using purpose written programs in Labview or Python. Image capture on A310 started 10 seconds before applying the heat source and finished 10 seconds after the heat source was switched off. As Ti55FT did not have any means of triggering other than through detection of temperature changes, image capture started before the heating started, and ended after the heating ended, and at least a few images were captured before and after heating. The same was done for corresponding measurements using the Seek Thermal Compact XR and Xenics Gobi640.  Measurements were conducted for 200 seconds, at Lock-In frequencies of 0.01, 0.02, 0.025, 0.04, 0.05 and 0.1~Hz, for SC7500. The measurements were only conducted at 0.01~Hz for the other cameras. The power from the halogen lamps were also varied up to its maximum of 200~W, to understand the effects of heating on experimental results, as shown in Figure~\ref{fig:setup}.

Images were exported to ASCII files from SC7500, A310 and Ti55FT using Altair, ResearchIR and SmartView softwares respectively. ASCII files were written directly from the custom software and Xeneth for the Seek Thermal Compact XR and Xenics Gobi640 respectively. The starting times for A310 and Ti55FT were located manually by observing the temperature-time profiles of single pixels on the measurements. The same was automated for the Seek Thermal Compact XR and Xenics Gobi640, as the controller could be synchronised with the camera frames over a WiFi network. The starting times were entered into a Python routine, for these all cameras except the SC7500, as it did not require this parameter due to the hardware trigger. All cameras did not have constant frame rates, so measurements were done in an asynchronous fashion. Hence, based on the starting time of the measurements, a purpose written algorithm in Python was used on all the data to create synchronised images that were spaced equally in time, by using linear interpolation of data between two frames. These generated typical heating profiles as illustrated in Figure~\ref{fig:setup}.

After the synchronised images were created, Lock-In calculations were applied to the data. The processing equations are similar to those of Bauer \etal~\cite{Bauer2009EDFA}. Weighting factors $K_{0}$ and $K_{-90}$ were defined in Equations~(\ref{eqn:K0}) and (\ref{eqn:Km90}) respectively, and these weighting factors for Lock-In frequency of 0.01~Hz and measurement times of 200 seconds are shown in Figure~\ref{fig:setup}. The Lock-In frequency is defined as $f$, and time for the $n$th frame is defined as $t_{n}$.

\begin{center}
\begin{equation}
K_{0}\left( t_{n} \right)=sin\left(2 \pi f t_{n}\right)
\label{eqn:K0}
\end{equation}
\end{center}

\begin{center}
\begin{equation}
K_{-90}\left( t_{n} \right)=-cos\left(2 \pi f t_{n}\right)
\label{eqn:Km90}
\end{equation}
\end{center}

For every pixel in row $r$ and column $c$, the weighting factors were multiplied with the image signals $I_{n}$ based on Equations~(\ref{eqn:S0}) and (\ref{eqn:Sm90}), where $N$ is the number of images per measurement. These routines generated images $S_{0}$ and $S_{-90}$. These images were then used to calculate the Lock-In amplitude ($A$) and phase ($\Phi$) images via Equations~(\ref{eqn:amplitude}) and (\ref{eqn:phase}) respectively. Note that $\Delta t$ and $T$ represent the time between consecutive frames and total duration of measurement (200 seconds) respectively. This was implemented to create Lock-In amplitude data that was independent of the camera frame rates. Also, if the measured signal is a pure sine wave, the values of $A$ would correspond to half the amplitude of that sine wave. Note that these equations consist of summation functions, and it is possible to calculate the corresponding images during image acquisition as described by Schlangen \etal~\cite{Schlangen2010MR}. On a computer, as the images are acquired, a parallel process can be used to do the corresponding calculations via multithreading. Given that the processing for each camera pixel was the same, it would also be possible to use a Graphics Processing Unit (GPU) to process the image data in real time.

\begin{center}
\begin{eqnarray}
S_{0}\left( r,c \right) & = & \sum_{n=1}^{N}{I_{n}\left( r,c \right)K_{0}\left( t_{n} \right)} \nonumber\\
                                  & = & \sum_{n=1}^{N}{I_{n}\left( r,c \right)sin\left(2 \pi f t_{n}\right)}
\label{eqn:S0}
\end{eqnarray}
\end{center}

\begin{center}
\begin{eqnarray}
S_{-90}\left( r,c \right) & = & \sum_{n=1}^{N}{I_{n}\left( r,c \right)K_{-90}\left( t_{n} \right)} \nonumber\\
                                     & = & -\sum_{n=1}^{N}{I_{n}\left( r,c \right)cos\left(2 \pi f t_{n}\right)}
\label{eqn:Sm90}
\end{eqnarray}
\end{center}

\begin{center}
\begin{eqnarray}
A\left( r,c \right) & = & \frac{\Delta t}{T}\sqrt{\left(S_{0}\left( r,c \right)\right)^{2}+\left(S_{-90}\left( r,c \right)\right)^{2}}
\label{eqn:amplitude}
\end{eqnarray}
\end{center}

\begin{center}
\begin{eqnarray}
\Phi\left( r,c \right) & = & tan^{-1}\left( \frac{-S_{-90}\left( r,c \right)}{S_{0}\left( r,c \right)} \right)
\label{eqn:phase}
\end{eqnarray}
\end{center}

\begin{figure} 
\begin{center}
\includegraphics[width=\figwidth mm]{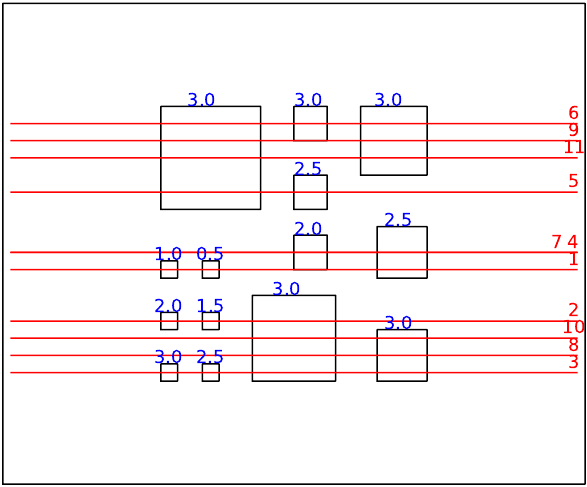}
\end{center}
\caption{Printed sample with square defects from 2~mm to 12~mm in width, each either 0.5~mm or 1~mm thick. Sample was 70~mm by 56~mm and 6~mm thick.  \textcolor{blue}{Blue} numbers above the defects indicate depth below the inspected surface and \textcolor{red}{red} horizontal lines with numbers indicate the numbered profiles where data was extracted from measurements.
\label{fig:printedsample2}}
\end{figure}

The samples with known defects (air pockets) were created using FDM 3D printers. The Makerbot Replicator 2 was used to print PLA samples, while the Makerbot Replicator Dual was used to print ABS samples. These samples were drawn using FreeCAD, and exported to STL files. The STL files were processed using Makerware to generate machine instructions for 3D printing these samples. There were two geometries that were created for the current study, as shown in Figures~\ref{fig:printedsample1} and \ref{fig:printedsample2}. The former consisted of 9 square defects (air pockets) 4~mm in width and 0.5~mm thick, spaced 10~mm apart in a 3 by 3 matrix. This sample was 70~mm by 56~mm by 8~mm, printed with 100\% infill, and defects were located 0.5~mm deep, in steps of 0.5~mm, to 4.5~mm. The latter consisted of more square defects (air pockets) of different sizes, from 2~mm in width, in steps of 2~mm, to 12~mm in width. Defects were 0.5~mm thick. This sample was 70~mm by 56~mm by 6~mm, printed with 100\% infill, and defects were located 0.5~mm deep, in steps of 0.5~mm, to 3.0~mm. Similar samples were printed, where defects were 1.0~mm thick. Such samples were printed in different orientations to investigate possible effects of print direction on bulk properties.

A purpose written Python algorithm was used to map the pixels on the phase images to real space, so as to locate the signals at the defects in a precise manner. Subsequently, data was exported along profiles as shown in Figures~\ref{fig:printedsample1} and \ref{fig:printedsample2}, to obtain quantitative data along rows of defects. Such quantitative data resulted in a better understanding of the characteristics of Lock-In phase signals with respect to the many parameters studied in the current scope.

\section{Results and Discussion}
\label{sect:results}

\begin{figure} 
\begin{center}
\includegraphics[width=\figwidth mm]{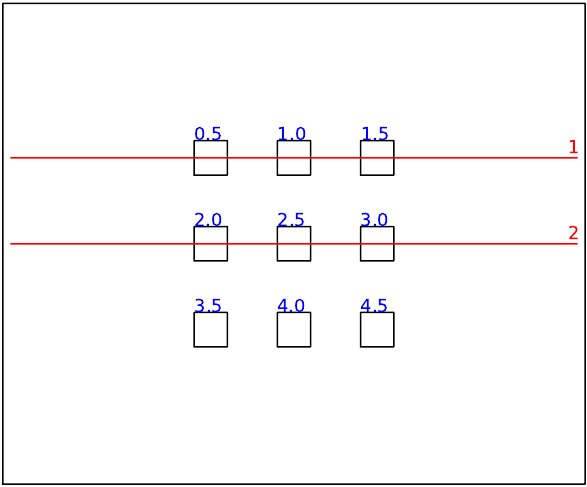}
\end{center}
\caption{Printed sample with 9 defects 4~mm by 4~mm in size, each 0.5~mm thick, spaced 10~mm apart in both directions. Sample was 70~mm by 56~mm and 8~mm thick. \textcolor{blue}{Blue} numbers above the defects indicate depth below the inspected surface and \textcolor{red}{red} horizontal lines with numbers indicate the numbered profiles where data was extracted from measurements.
\label{fig:printedsample1}}
\end{figure}

Many variables were explored in the current study, to understand their relative effects on the results. The reference measurements were done using SC7500, at 1 frame per second, for 200 seconds, with Lock-In frequency of 0.01~Hz at low power, on PLA samples with defects of 4~mm in width (see Figure~\ref{fig:printedsample1}). Unless otherwise stated, the measurement parameters presented refer to those of the reference measurements.

\begin{figure} 
\begin{center}
\includegraphics[width=\figwidth mm]{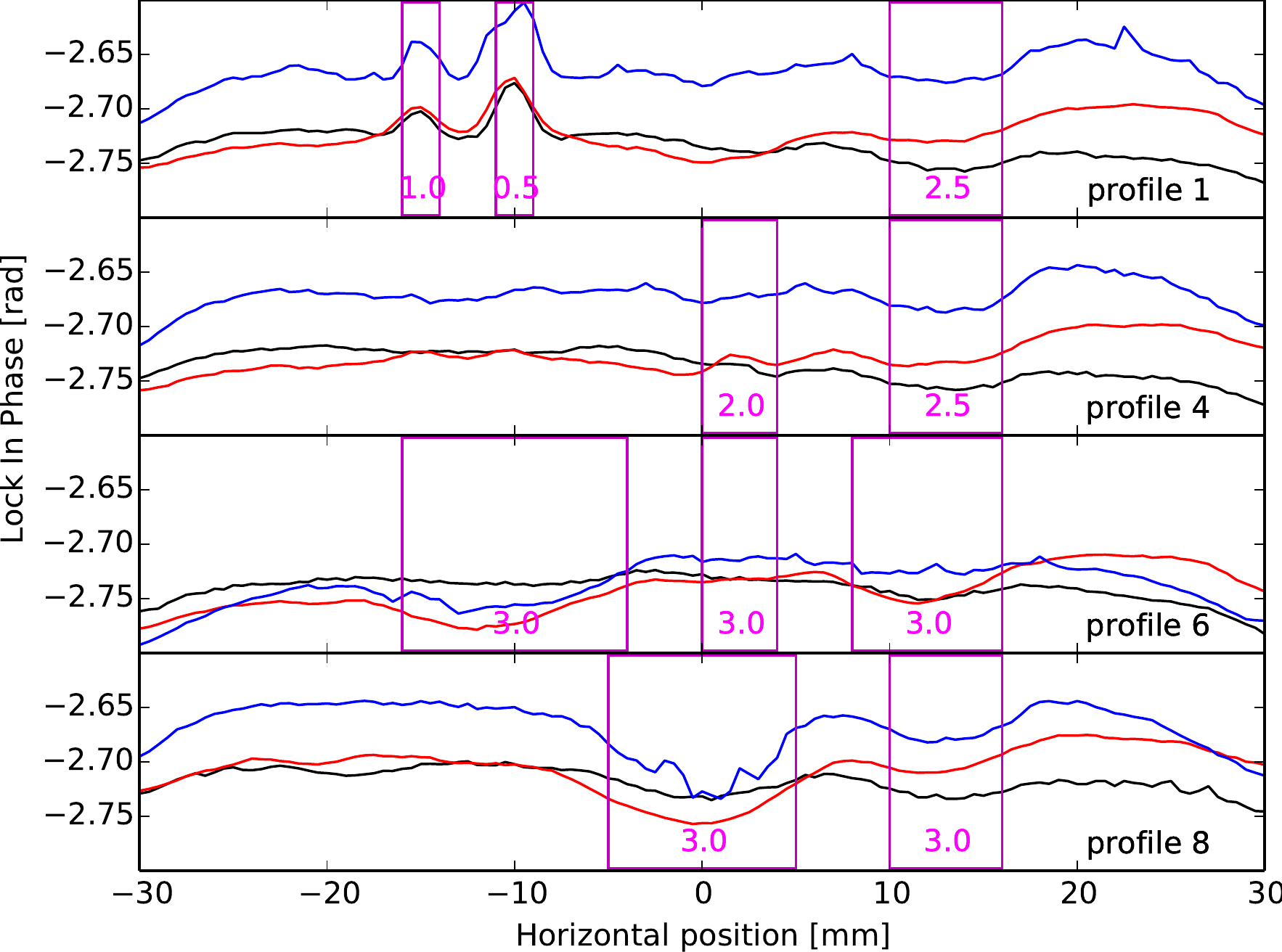}
\end{center}
\caption{Lock-in phase signals from measurements conducted at 0.01~Hz lock-in frequency, at 1~frame per second, for 200~seconds, for PLA samples shown in Figure~\ref{fig:printedsample2}. \textcolor{black}{Black} lines represent data extracted from a sample with defects that were 0.5~mm thick. \textcolor{red}{Red} and \textcolor{blue}{blue} lines represent data extracted from vertically and horizontally printed samples with defects 1.0~mm thick. Data is presented for profiles 1, 4, 6, and 8. \textcolor{magenta}{Magenta} boxes  represent the locations of known defects, with numbers indicating the depth of defects.
\label{fig:phaseprintdir}}
\end{figure}

\begin{figure} 
\begin{center}
\includegraphics[width=\figwidth mm]{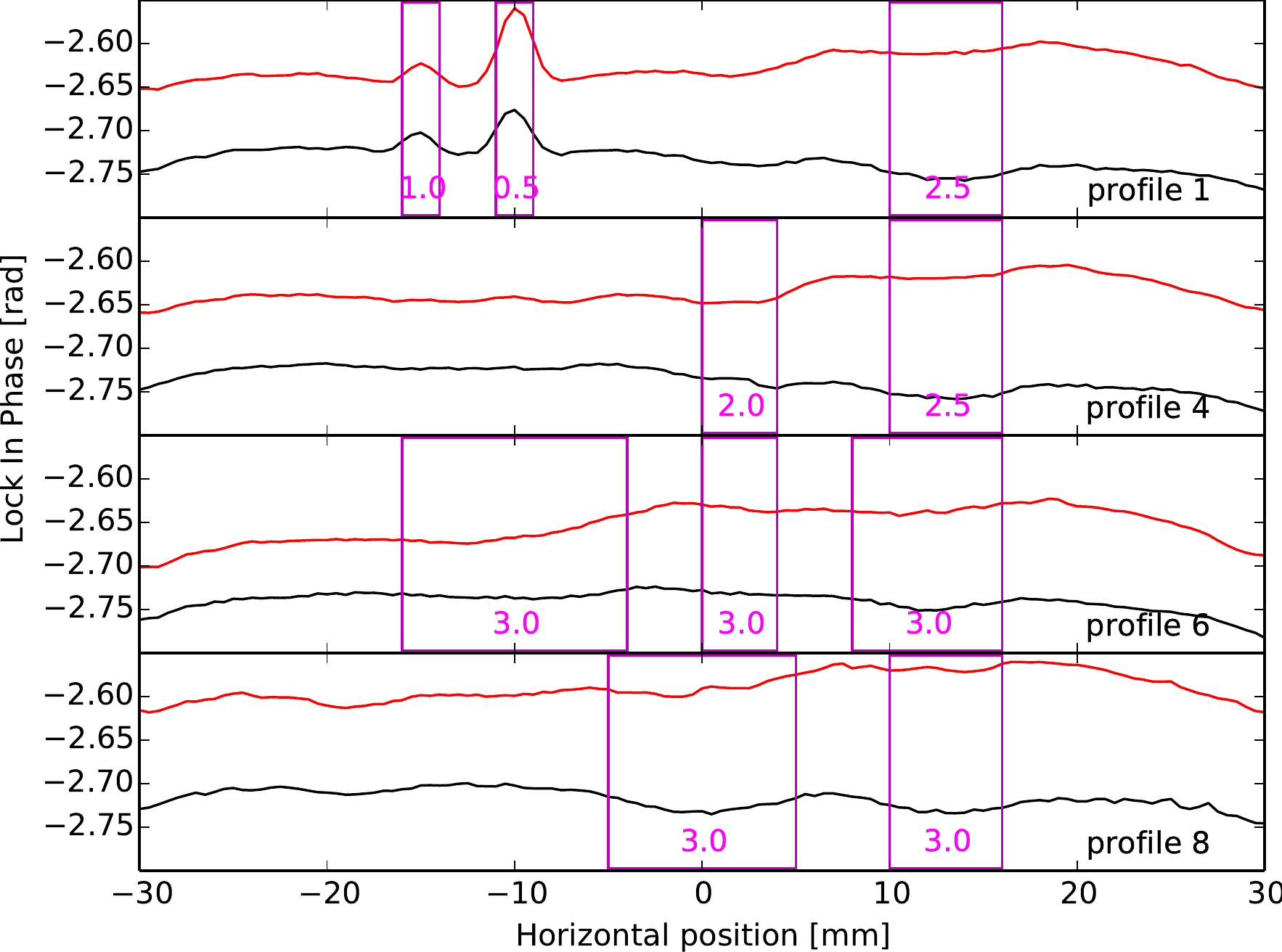}
\end{center}
\caption{Lock-in phase signals from measurements conducted at 0.01~Hz lock-in frequency, at 1~frame per second, for 200~seconds, for samples shown in Figure~\ref{fig:printedsample2}. \textcolor{black}{Black} and \textcolor{red}{red} lines represent data extracted from PLA and ABS samples with defects that were 0.5~mm thick. Data is presented for profiles 1, 4, 6, and 8. \textcolor{magenta}{Magenta} boxes  represent the locations of known defects, with numbers indicating the depth of defects.
\label{fig:phasematsqdiff}}
\end{figure}

\begin{figure} 
\begin{center}
\includegraphics[width=\figwidth mm]{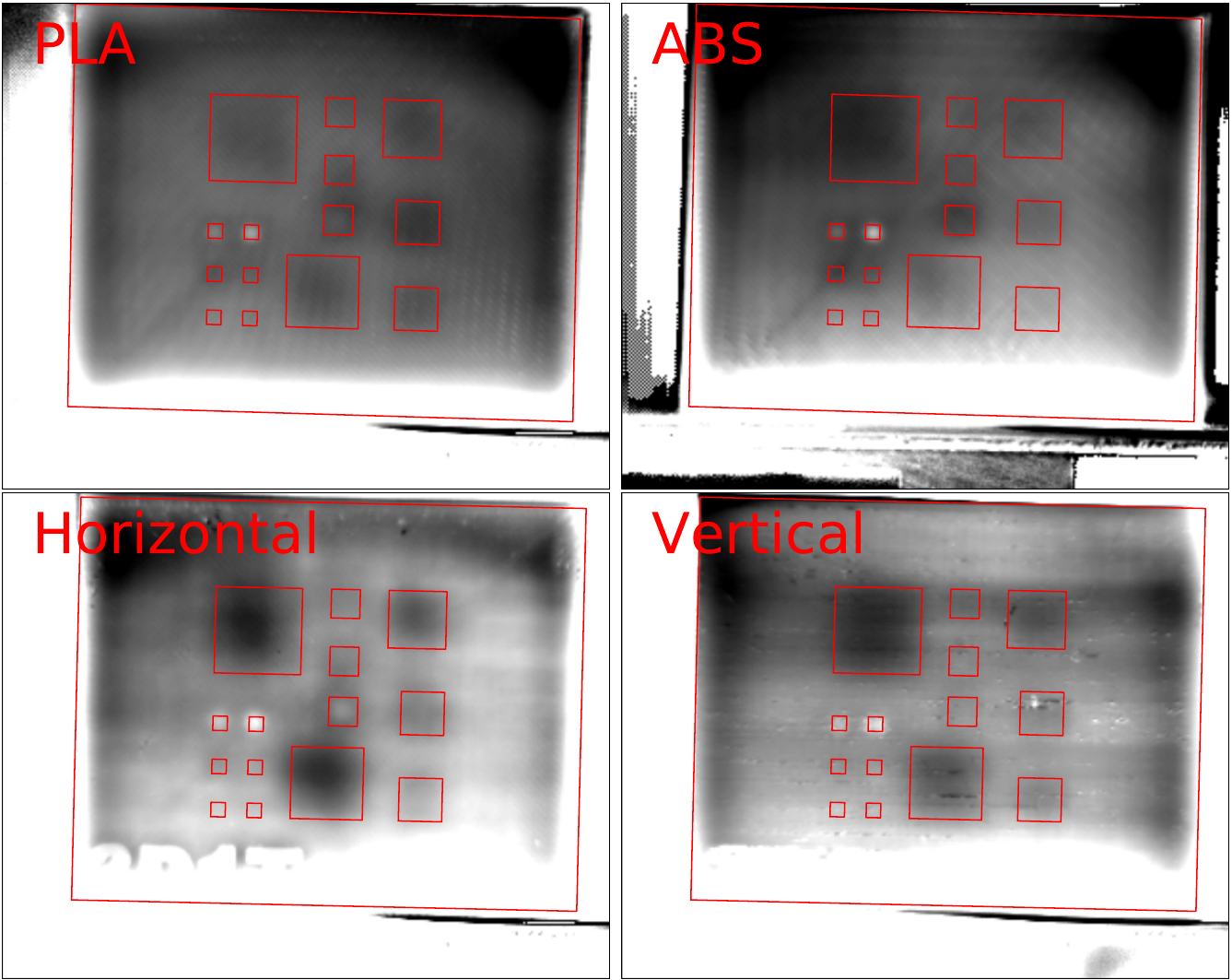}
\end{center}
\caption{Lock-in phase images from measurements conducted at 0.01~Hz lock-in frequency, at 1~frame per second, for 200~seconds, for samples shown in Figure~\ref{fig:printedsample2}. Top row - PLA (left) and ABS (right) samples with defects that were 0.5~mm thick. Bottom row - PLA samples that were printed horizontally (left) and vertically (right) with defects 1.0~mm thick.
\label{fig:phaseimgsqdiff}}
\end{figure}

Figure~\ref{fig:phaseprintdir} compares print direction and defect thickness, as well as the relative effects of defect width on the phase signals. Only profiles 1, 4 6, and 8 were shown to illustrate the relative effects of these variables. Horizontally printed samples were printed with the largest face on the print platform, while vertically printed samples were printed with the samples standing on the face with the width and thickness edges on the print platform. As shown by the data, the print direction appeared to result in a phase shift. However, this did not appear to affect the relative phase contrast resulting from defects. The increase in thickness of defects from 0.5~mm to 1.0~mm resulted in stronger signal contrast, especially for deeper defects at 3.0~mm below the surface of the sample, as shown by profiles 6 and 8. Results show that sub-surface defects as small as 2~mm in width can be detected at 0.5 and 1.0~mm depths, with clear peaks. Defects of width 4~mm can be observed up to around 2.0~mm with signals as weak troughs. Defects of width 6~mm were visible even at 3~mm depth, with greater contrast when the thickness of the defects increased. As defects increased in width to 8~mm, 10~mm, and 12~mm, signal contrast was more prominent. Qualitatively, it appears that the width of the defects have to be at least around twice their corresponding depth in order to be detected. Figure~\ref{fig:phaseimgsqdiff} shows the corresponding phase images for these measurements, and these trends can also be observed.
Figure~\ref{fig:phasematsqdiff} compares ABS and PLA with different sized defects. The relative contrast due to defects were similar between the two materials, despite a clear phase shift. This indicated a stronger dependence on geometry of defects. 

\begin{figure} 
\begin{center}
\includegraphics[width=\figwidth mm]{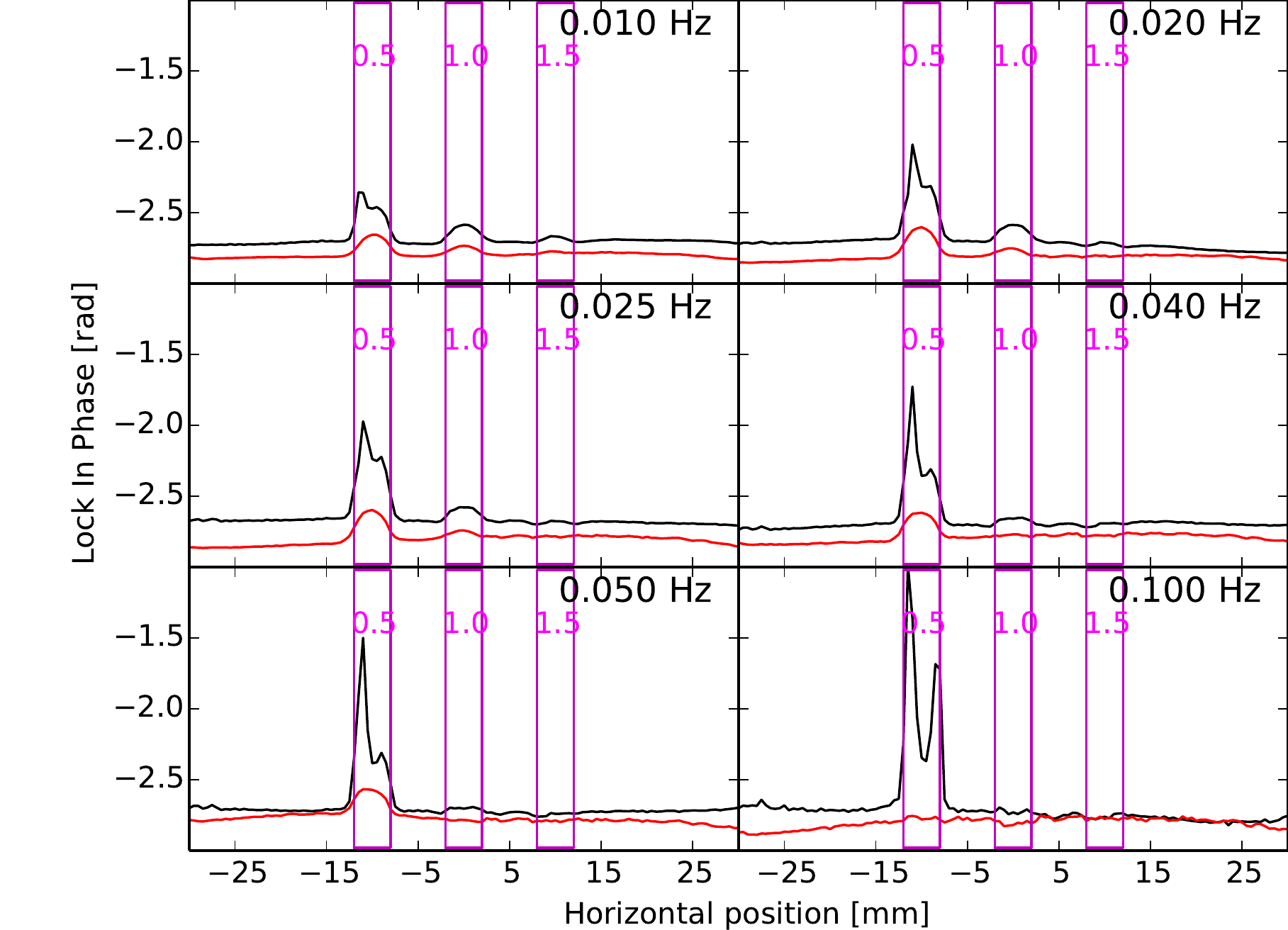}
\end{center}
\caption{Lock-in phase signals from measurements conducted at different lock-in frequencies, at 1~frame per second, for 200~seconds, for samples shown in Figure~\ref{fig:printedsample1}. \textcolor{black}{Black} and \textcolor{red}{red} lines represent data extracted from PLA and ABS samples with defects that were 0.5~mm thick. Data is presented for profile 1. \textcolor{magenta}{Magenta} boxes  represent the locations of known defects, with numbers indicating the depth of defects.
\label{fig:phasematfreqprof1}}
\end{figure}

\begin{figure} 
\begin{center}
\includegraphics[width=\figwidth mm]{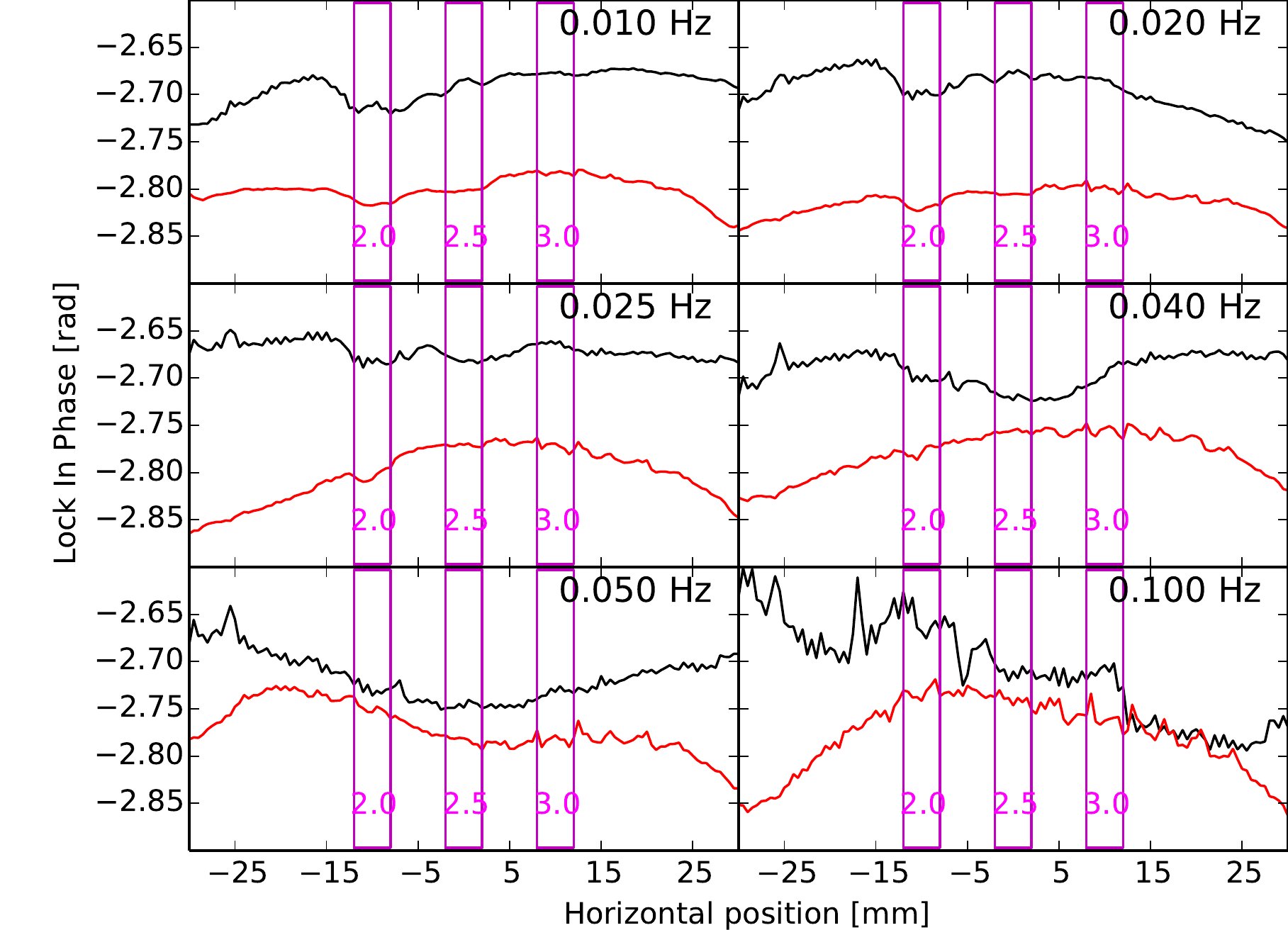}
\end{center}
\caption{Lock-in phase signals from measurements conducted at different lock-in frequencies, at 1~frame per second, for 200~seconds, for samples shown in Figure~\ref{fig:printedsample1}. \textcolor{black}{Black} and \textcolor{red}{red} lines represent data extracted from PLA and ABS samples with defects that were 0.5~mm thick. Data is presented for profile 2. \textcolor{magenta}{Magenta} boxes  represent the locations of known defects, with numbers indicating the depth of defects.
\label{fig:phasematfreqprof2}}
\end{figure}

\begin{figure} 
\begin{center}
\includegraphics[width=\figwidth mm]{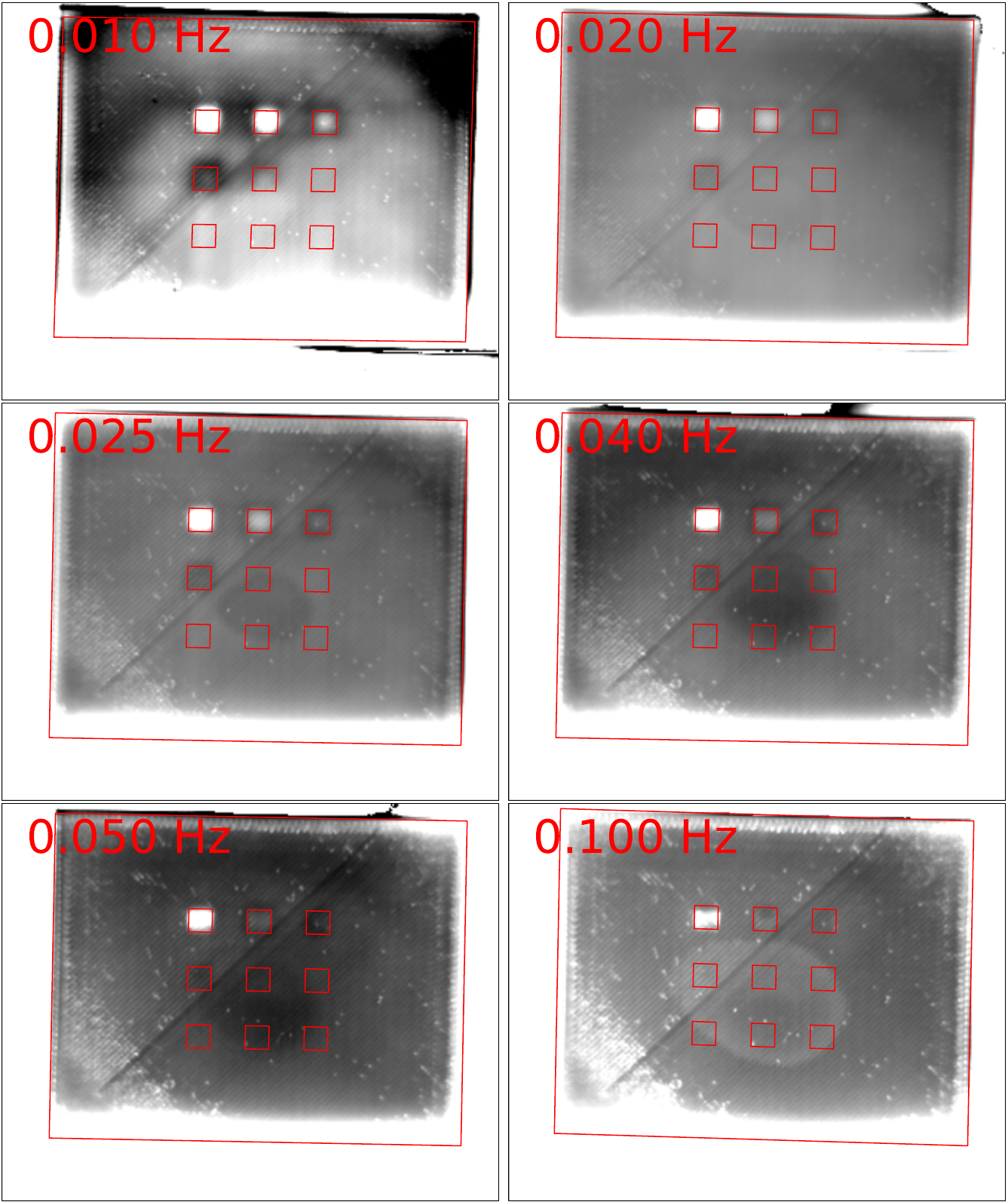}
\end{center}
\caption{Lock-in phase images from measurements conducted at different lock-in frequencies, at 1~frame per second, for 200~seconds, for samples shown in Figure~\ref{fig:printedsample1}, printed using PLA.
\label{fig:phaseimgfreqPLA}}
\end{figure}

\begin{figure} 
\begin{center}
\includegraphics[width=\figwidth mm]{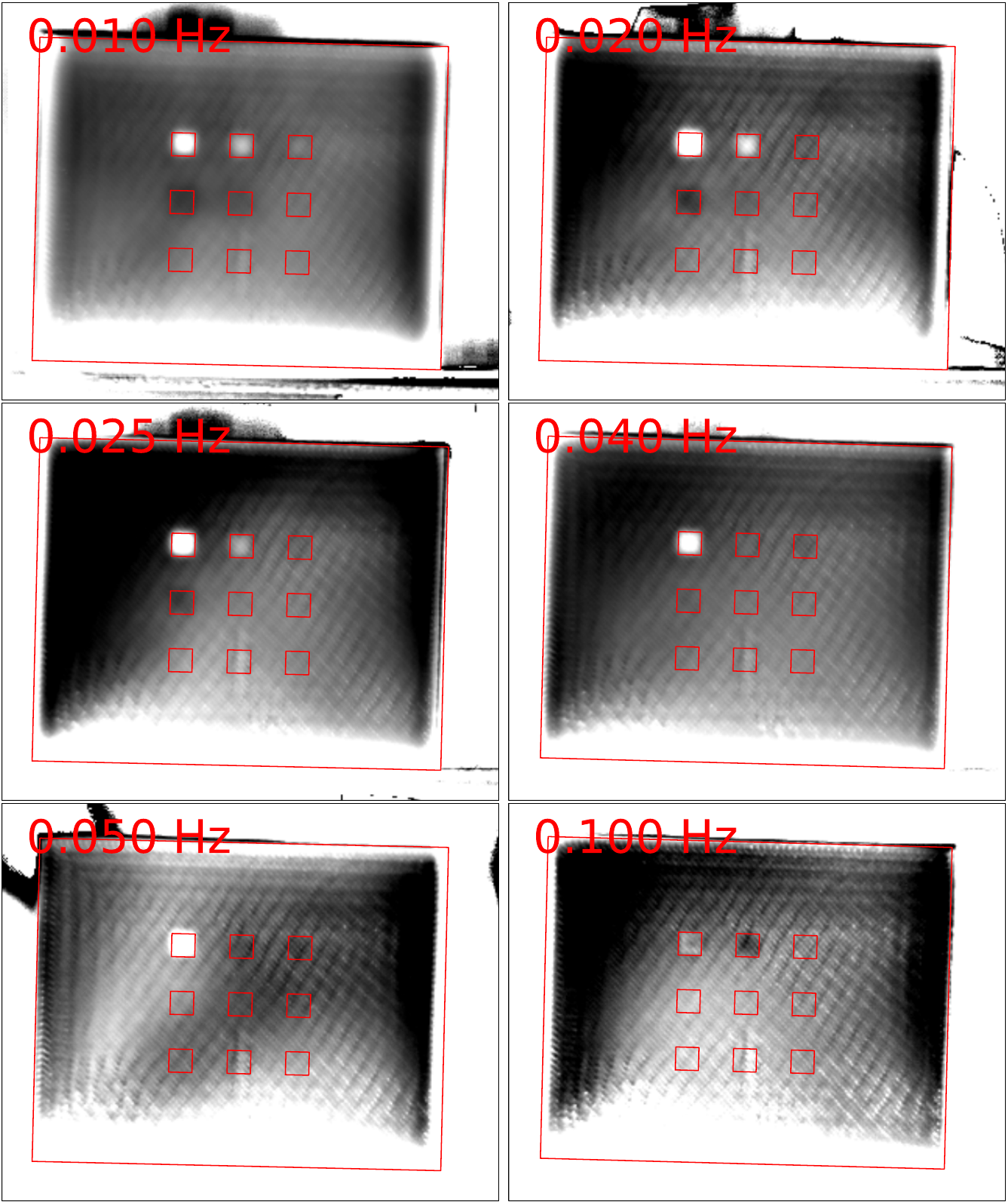}
\end{center}
\caption{Lock-in phase images from measurements conducted at different lock-in frequencies, at 1~frame per second, for 200~seconds, for samples shown in Figure~\ref{fig:printedsample1}, printed using ABS.
\label{fig:phaseimgfreqABS}}
\end{figure}

\begin{figure} 
\begin{center}
\includegraphics[width=\figwidth mm]{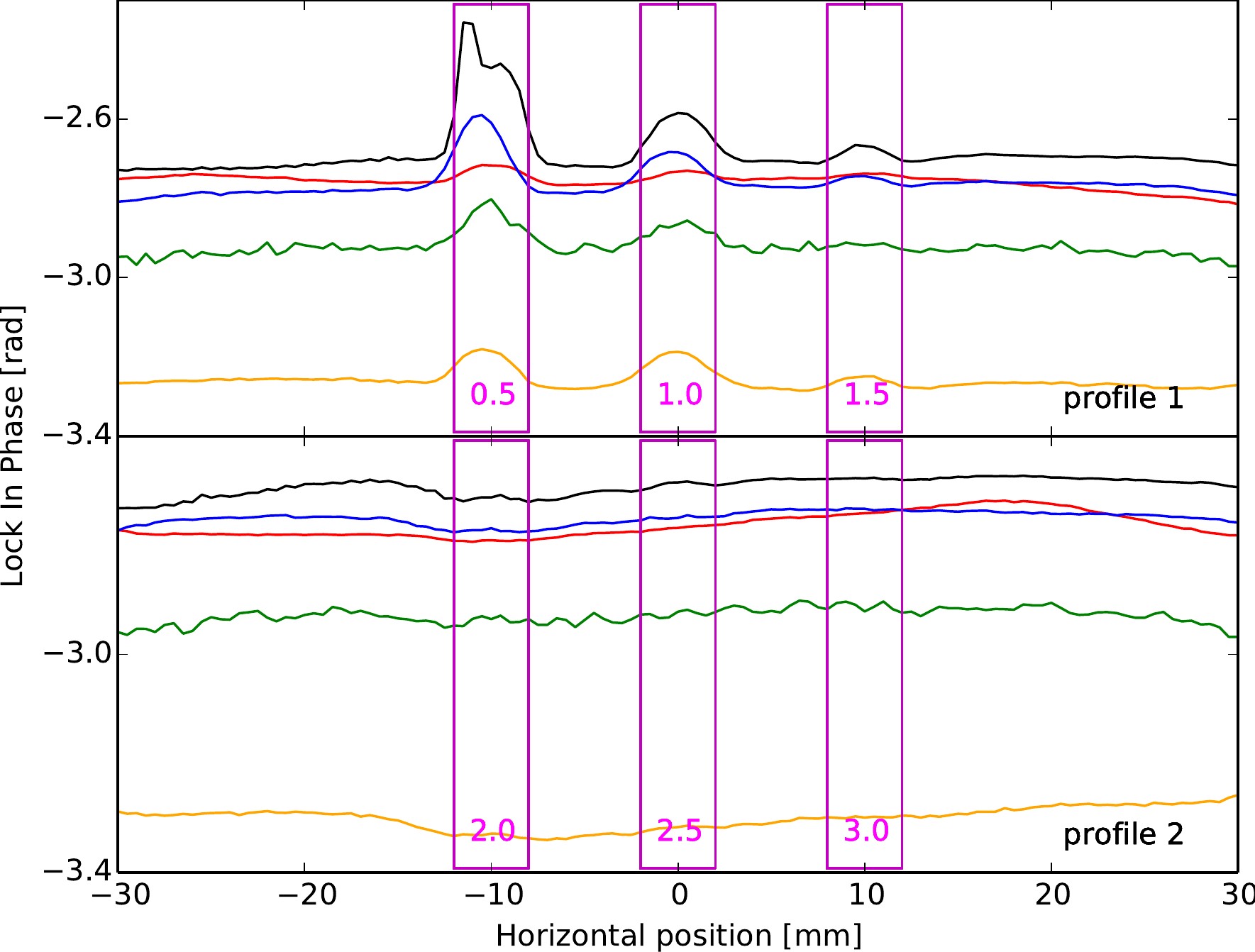}
\end{center}
\caption{Lock-in phase signals from measurements conducted at 0.01~Hz lock-in frequency, for 200~seconds, for PLA samples shown in Figure~\ref{fig:printedsample1}. \textcolor{black}{Black}, \textcolor{red}{red}, \textcolor{blue}{blue}, \textcolor{OliveGreen}{green}, and \textcolor{Dandelion}{orange} lines represent data extracted from using SC7500 at 1 frame per second, A310 at 2 frames per second, Ti55FT at 0.5 frames per second, Seek Thermal CompactXR at 1 frame per second and Gobi 640 at 1 frame per second respectively. Data is presented for profiles 1 and 2. \textcolor{magenta}{Magenta} boxes  represent the locations of known defects, with numbers indicating the depth of defects.
\label{fig:phasematcam}}
\end{figure}

Figures~\ref{fig:phasematfreqprof1} and \ref{fig:phasematfreqprof2} show the Lock-In phase signals for ABS and PLA samples with defects 4~mm in width at different depths, along profiles 1 and 2 respectively. The geometry of these samples is shown in Figure~\ref{fig:printedsample1}. It is shown clearly that there exists strong peaks for the PLA sample for the defect at 0.5~mm depth. However, this could be due to the swelling of the defect at shallow depths. The swelling of the defect resulted in more reflections on the surface, which could have resulted in a spurious large contrast as observed. The SC7500 was the only camera where this is observed, as it is sensitive in the Mid Wave Infrared (MWIR) range, and the heat source emitted significant radiation in that range. In contrast, the other cameras did not pick up reflected signals as they were only sensitive in the Long Wave Infrared (LWIR) range (see Figure~\ref{fig:phasematcam}). This is further substantiated by the clear double peaks for the defect at 0.5~mm depth in the PLA sample, at all Lock-In frequencies, which is only possible due to reflected signals. By comparing other defects between both materials, the trends in relative contrast can be said to be similar. As shown in these figures, the defects at 1.0~mm depth started to disappear when Lock-In frequencies were increased to 0.1~Hz. Corresponding trends for the defects that were 2.0~mm and 2.5~mm deep appeared at around 0.04~Hz and 0.01~Hz respectively. These results showed that as Lock-In frequency decreased, the signal contrast increased, and defects of width 4.0~mm could be detected up to around 2.5~mm depth. Hence, it may be possible to have qualitative detection of defects by simply using the lowest possible Lock-In frequencies, subject to time and equipment constraints.
Figures~\ref{fig:phaseimgfreqPLA} and \ref{fig:phaseimgfreqABS} show the corresponding phase images for PLA and ABS at different lock in frequencies, and it can also be observed that the signal contrast is enhanced at lower frequencies, and deeper defects can only be detected at lower Lock-In frequencies, similar to the findings of Duan \etal~\cite{Duan2013IPT}.

\begin{figure} 
\begin{center}
\includegraphics[width=\figwidth mm]{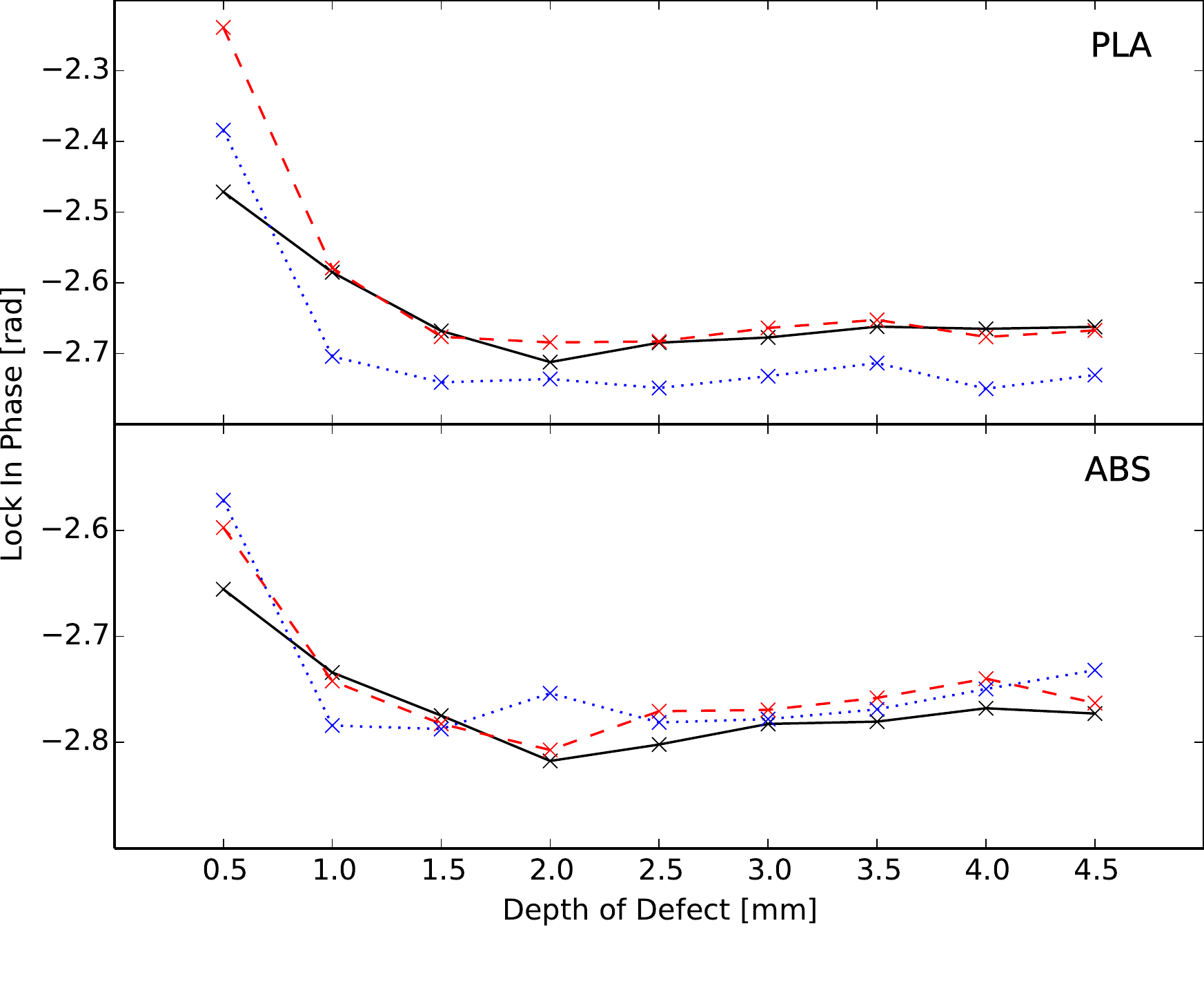}
\end{center}
\caption{Lock-in phase signals from measurements conducted at different lock-in frequencies, at 1~frame per second, for 200~seconds, for samples shown in Figure~\ref{fig:printedsample1}, for PLA~(top) and ABS~(bottom) samples with defects that were 0.5~mm thick. Signals from centroid of the 9 defects were used to obtain phase signals with respect to depth, at lock-in frequencies of 0.01~Hz (\textcolor{black}{$\times$, solid}), 0.025~Hz (\textcolor{red}{$\times$, dashed}), and 0.05~Hz (\textcolor{blue}{$\times$, dotted}).
\label{fig:depthresolution}}
\end{figure}

To further illustrate the relative effects of depth of defects and Lock-In frequency on the relative contrast, Lock-In phase data was obtained from the centroids of defects for both ABS and PLA. These results are shown in Figure~\ref{fig:depthresolution} for Lock-In frequencies of 0.01, 0.025 and 0.05~Hz. The trends show that deeper defects can only be detected by lowering the Lock-In frequency. Also, it is interesting to note that for Lock-In frequency of 0.01~Hz, data from both materials show a decrease for defects up to 2.0~mm depth, and a slight increase towards asymptotic values. These results indicate that automated detection of defects is trivial only up to around 1.5~mm depth for the samples used in the current study, as relative phase contrast is higher, and the Lock-In phase signals would only point to single values of depth of defects. For defects deeper than 1.5~mm beneath the sample surface, different depths of defects result in similar values of Lock-In phase, so single measurements of a sample would not result in automated defect detection for these depths, and it may only be possible via visual inspection of the phase images. An approach for automated defect detection is shown by Spiessberger \etal~\cite{Spiessberger2010NDTA}, where calibration tables were used to determine the thickness of a polymer wedge sample using Lock-In thermography.

In the current study, five different cameras were used to detect defects in PLA samples with defects 4.0~mm in width (see Figure~\ref{fig:printedsample1}). The Lock-In phase signals for profiles 1 and 2 are shown in Figure~\ref{fig:phasematcam}. It can be observed that there appears to be minimal differences in the relative phase contrast at each defect, except the profiles measured using the A310. This could be due to the lower heating power used for that measurement. The heating power was the same for all other cameras used in this comparison. In addition, the baseline phase values were different for each camera, due to differences in frame rates and units of measurements. It was also observed that the SC7500, which is the only camera sensitive in the Mid Wave Infrared (MWIR) wavelength range, was the only camera able to detect the reflected heat signals from the halogen lamps. Nevertheless, it is clearly shown that Lock-In thermography can be conducted using different cameras with minimal differences in phase contrast. Hence, it is possible to use cheap equipment (e.g. Seek Thermal CompactXR) in Lock-In thermography to detect defects in samples, unlike other techniques known to date. 

\begin{figure} 
\begin{center}
\includegraphics[width=\figwidth mm]{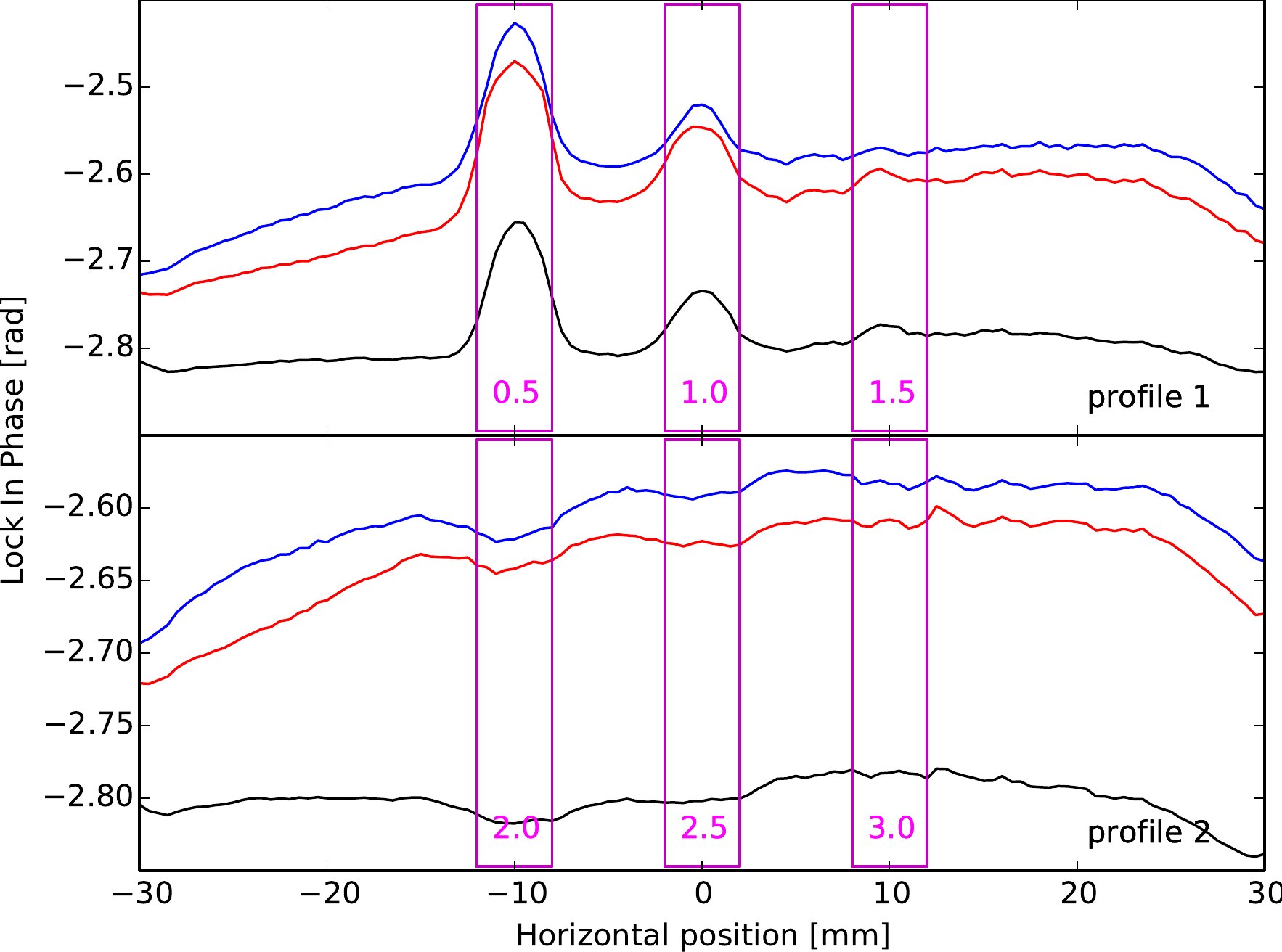}
\end{center}
\caption{Lock-in phase signals from measurements conducted at 0.01~Hz lock-in frequency, for 200~seconds, for ABS samples shown in Figure~\ref{fig:printedsample1}. \textcolor{black}{Black}, \textcolor{red}{red} and \textcolor{blue}{blue} lines represent data for low, medium and high power. Data is presented for profiles 1 and 2. \textcolor{magenta}{Magenta} boxes  represent the locations of known defects, with numbers indicating the depth of defects.
\label{fig:phasematpow}}
\end{figure}

\begin{figure} 
\begin{center}
\includegraphics[width=\figwidth mm]{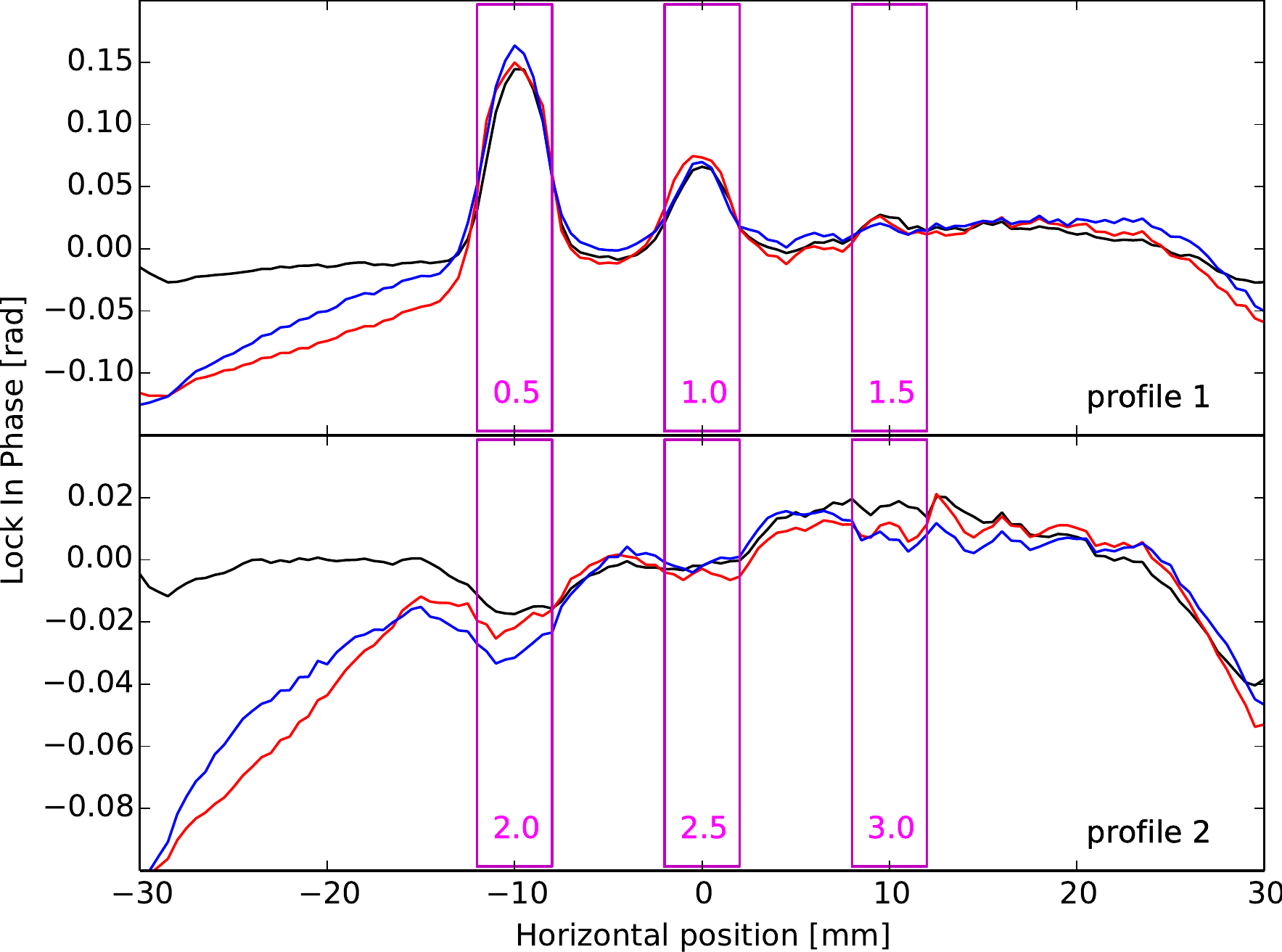}
\end{center}
\caption{Lock-in phase signals from measurements conducted at 0.01~Hz lock-in frequency, for 200~seconds, for ABS samples shown in Figure~\ref{fig:printedsample1}. \textcolor{black}{Black}, \textcolor{red}{red} and \textcolor{blue}{blue} lines represent data with baseline correction for low, medium and high power. Data is presented for profiles 1 and 2. \textcolor{magenta}{Magenta} boxes  represent the locations of known defects, with numbers indicating the depth of defects.
\label{fig:phasematpowoffset}}
\end{figure}

Different power settings were used to investigate the effects of heating power on the Lock-In phase. These results were obtained using ABS samples, with signal profiles 1 and 2 in Figure~\ref{fig:phasematpow}. Baseline offset was applied to compare the relative contrast, as shown in Figure~\ref{fig:phasematpowoffset}. As shown in Figure~\ref{fig:phasematpow}, the Lock-In phase increased as the power increased, as shown in the trends for the baseline. However, there were no apparent changes in the relative contrast when the heating power changed, as shown in Figure~\ref{fig:phasematpowoffset}. The disparities in the shifted profiles on the left side of Figure~\ref{fig:phasematpowoffset} could be due to the changes in positions of some of the halogen lamps. These results show that Lock-In thermography is not sensitive to heating parameters, and reinforces the hypothesis that geometry is the predominant factor in influencing these signals.

\begin{figure} 
\begin{center}
\includegraphics[width=\figwidth mm]{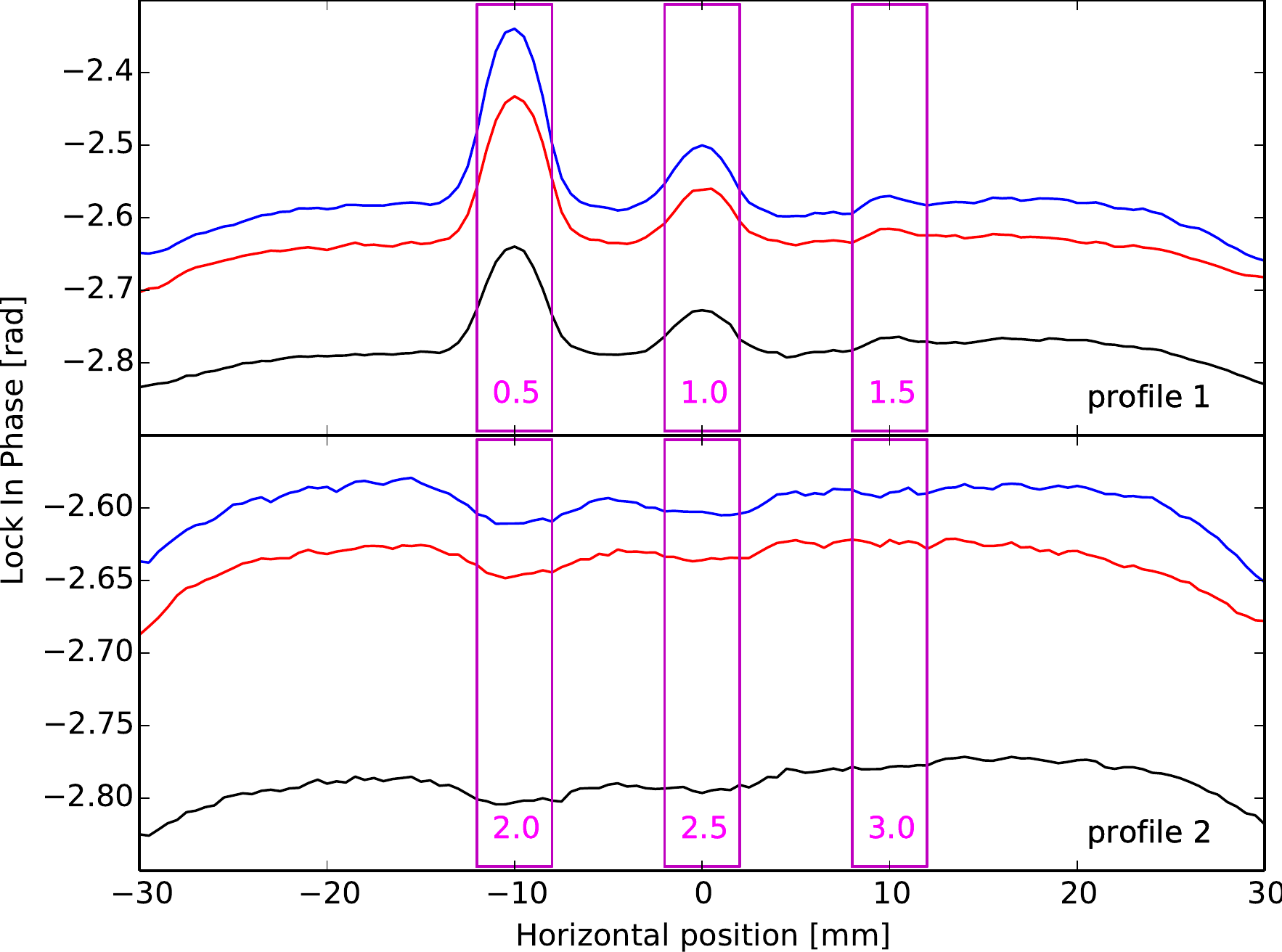}
\end{center}
\caption{Lock-in phase signals from measurements conducted at 0.01~Hz lock-in frequency, for 200~seconds, for ABS samples shown in Figure~\ref{fig:printedsample1}, using Ti55FT. \textcolor{black}{Black}, \textcolor{red}{red} and \textcolor{blue}{blue} lines represent data for low, medium and high power. Data is presented for profiles 1 and 2. \textcolor{magenta}{Magenta} boxes  represent the locations of known defects, with numbers indicating the depth of defects.
\label{fig:phasematpowTi55}}
\end{figure}

\begin{figure} 
\begin{center}
\includegraphics[width=\figwidth mm]{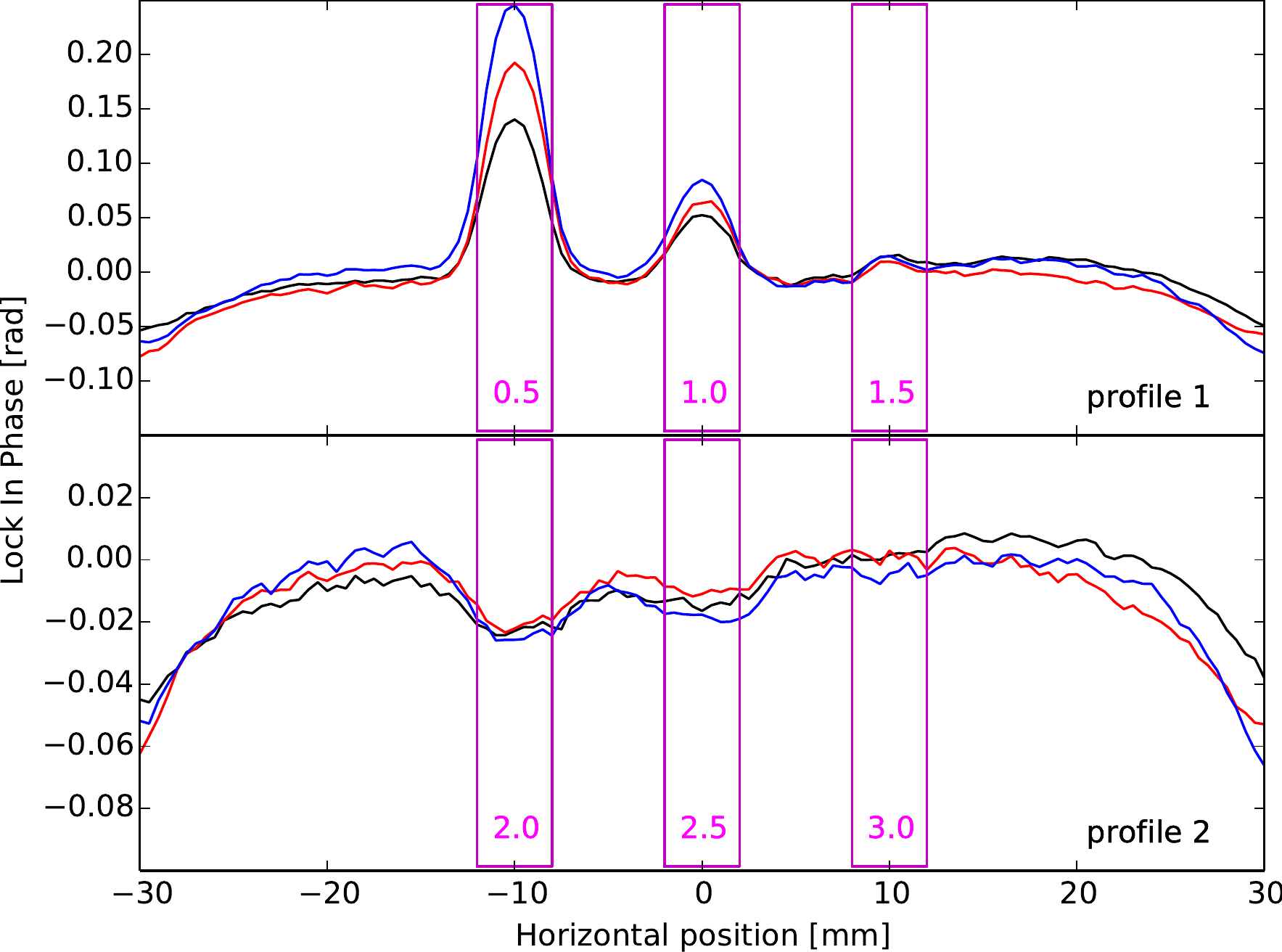}
\end{center}
\caption{Lock-in phase signals from measurements conducted at 0.01~Hz lock-in frequency, for 200~seconds, for ABS samples shown in Figure~\ref{fig:printedsample1}, using Ti55FT. \textcolor{black}{Black}, \textcolor{red}{red} and \textcolor{blue}{blue} lines represent data with baseline correction for low, medium and high power. Data is presented for profiles 1 and 2. \textcolor{magenta}{Magenta} boxes  represent the locations of known defects, with numbers indicating the depth of defects.
\label{fig:phasematpowoffsetTi55}}
\end{figure}

\begin{figure} 
\begin{center}
\includegraphics[width=\figwidthtwo mm]{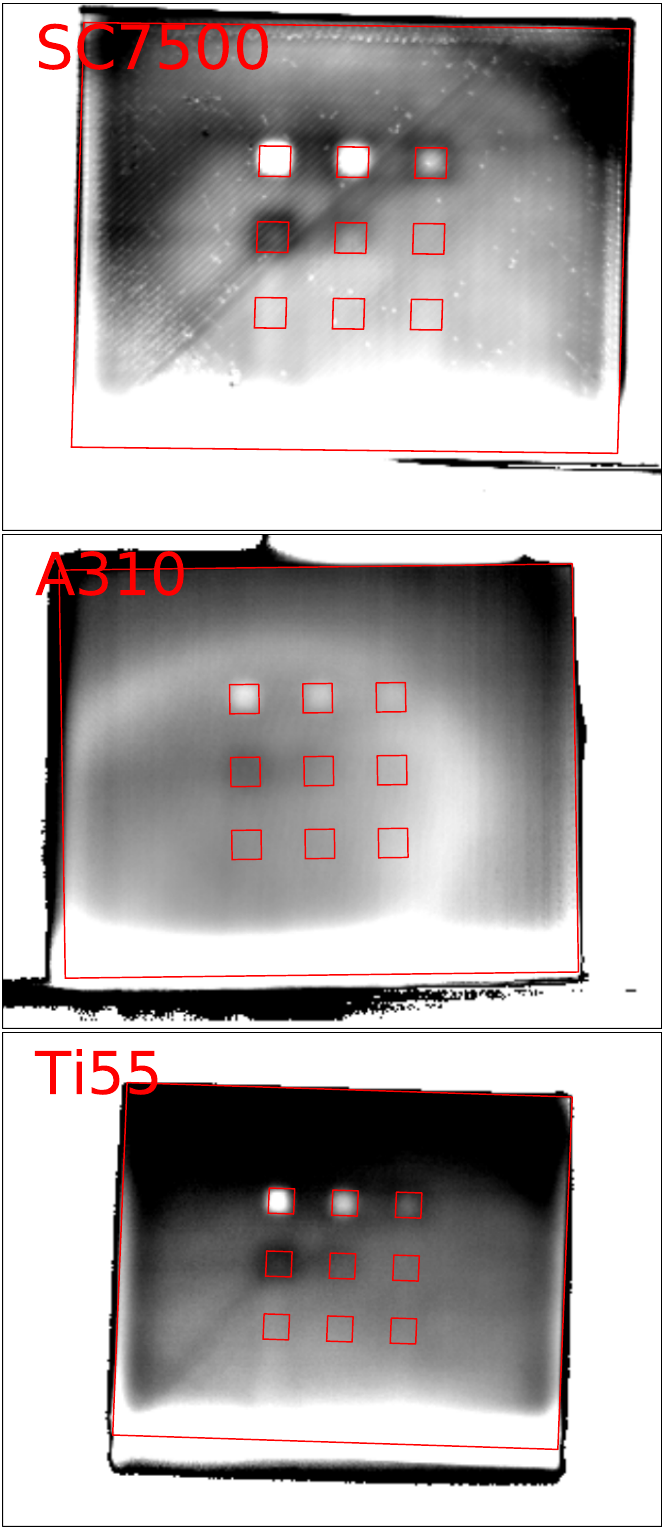}
\includegraphics[width=\figwidthtwo mm]{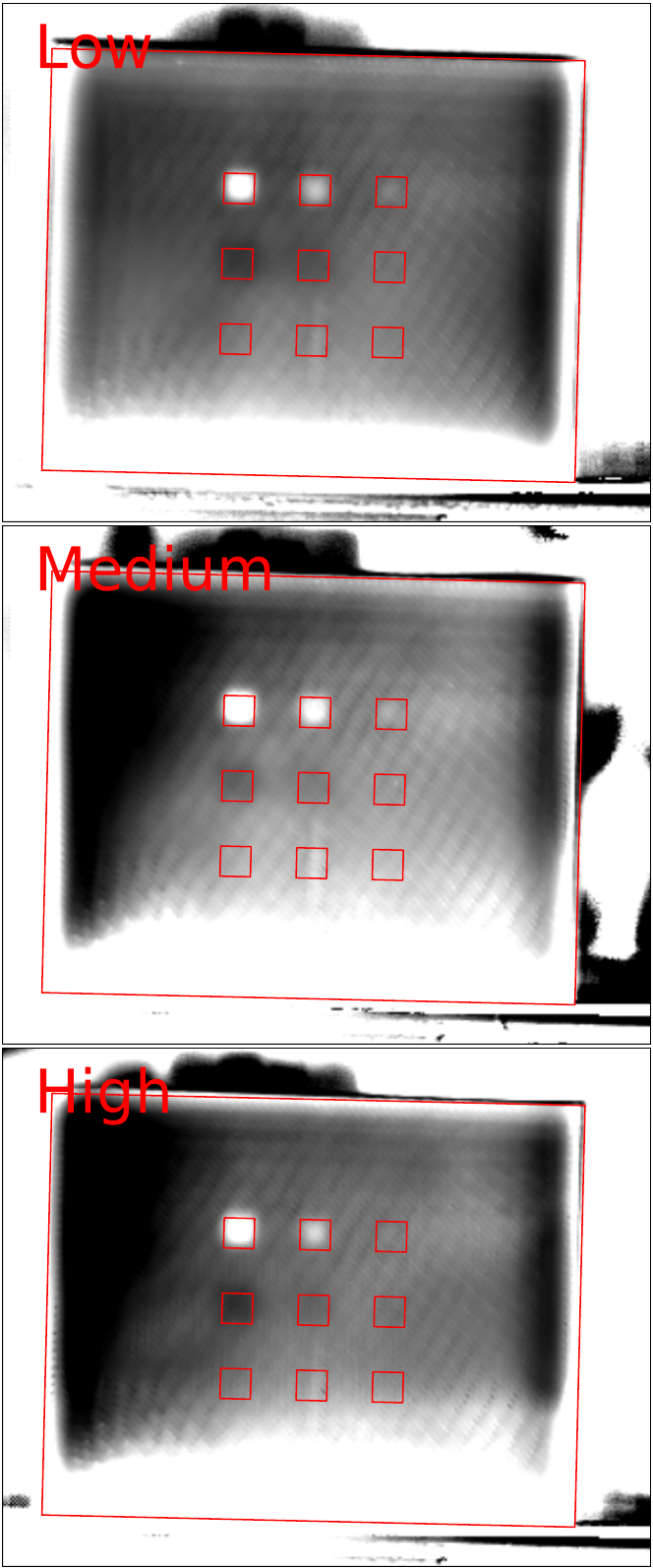}
\includegraphics[width=\figwidthtwo mm]{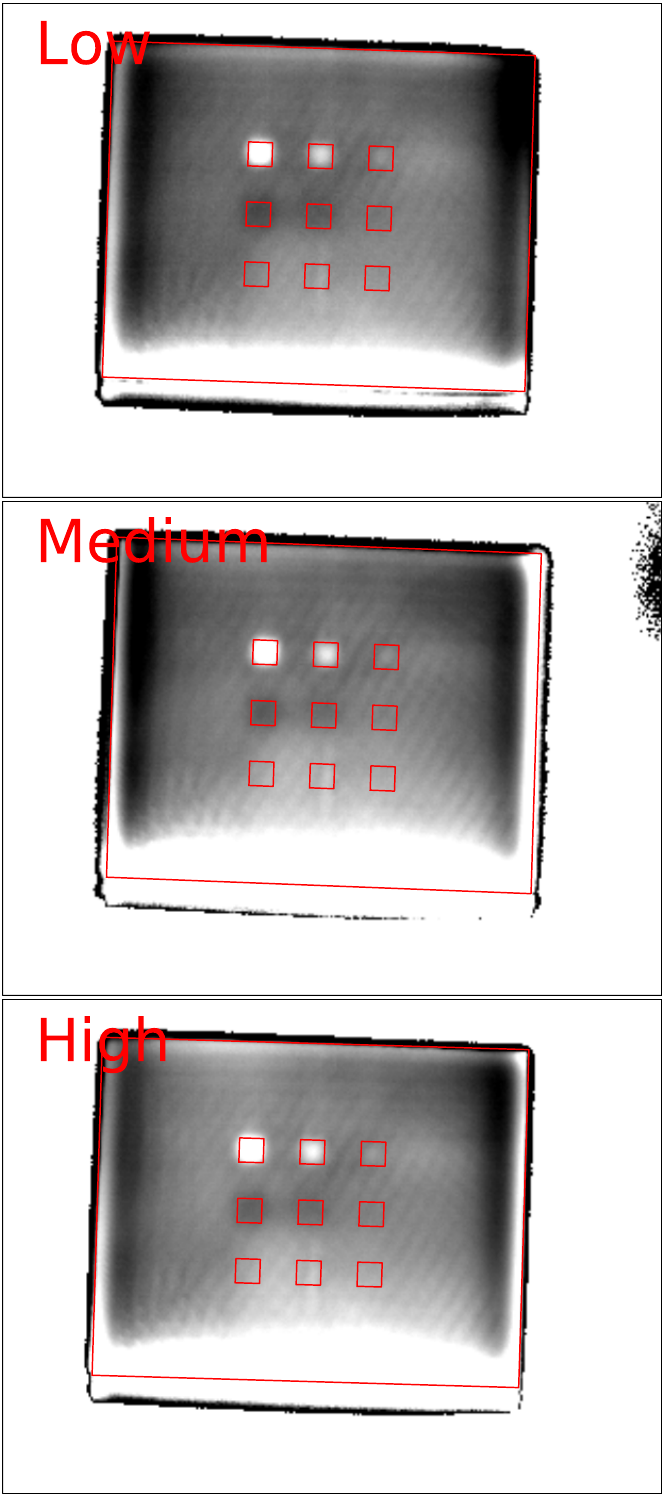}
\end{center}
\caption{Lock-in phase images from measurements conducted at 0.01~Hz lock-in frequency, for 200~seconds, for samples shown in Figure~\ref{fig:printedsample1}, printed using PLA (left) and ABS (middle and right). Left column: measurements were conducted on the SC7500 at 1 frame per second, A310 at 2 frames per second, and Ti55FT at 0.5 frames per second. Middle and right columns: measurements were conducted using low, medium and high power, using SC7500 (middle) and Ti55FT (right).
\label{fig:phaseimgCameraPower}}
\end{figure}

The same benchmarking was done using the Ti55FT, with corresponding results shown in Figures~\ref{fig:phasematpowTi55} and~\ref{fig:phasematpowoffsetTi55}. Similar baseline shifts were observed when heating power increased. However, it is interesting to note that when the heating power increased, the relative phase contrast at defects located 0.5~mm and 1.0~mm below the surface increased. This is an interesting phenomenon that may be better understood in future work.
The corresponding Lock-In phase images for different cameras and heating power are shown in Figure~\ref{fig:phaseimgCameraPower}. It can be observed that the A310 had a different heating setup with a less uniform heat source, as fewer halogen lamps were used for that measurement.

\begin{figure} 
\begin{center}
\includegraphics[width=\figwidth mm]{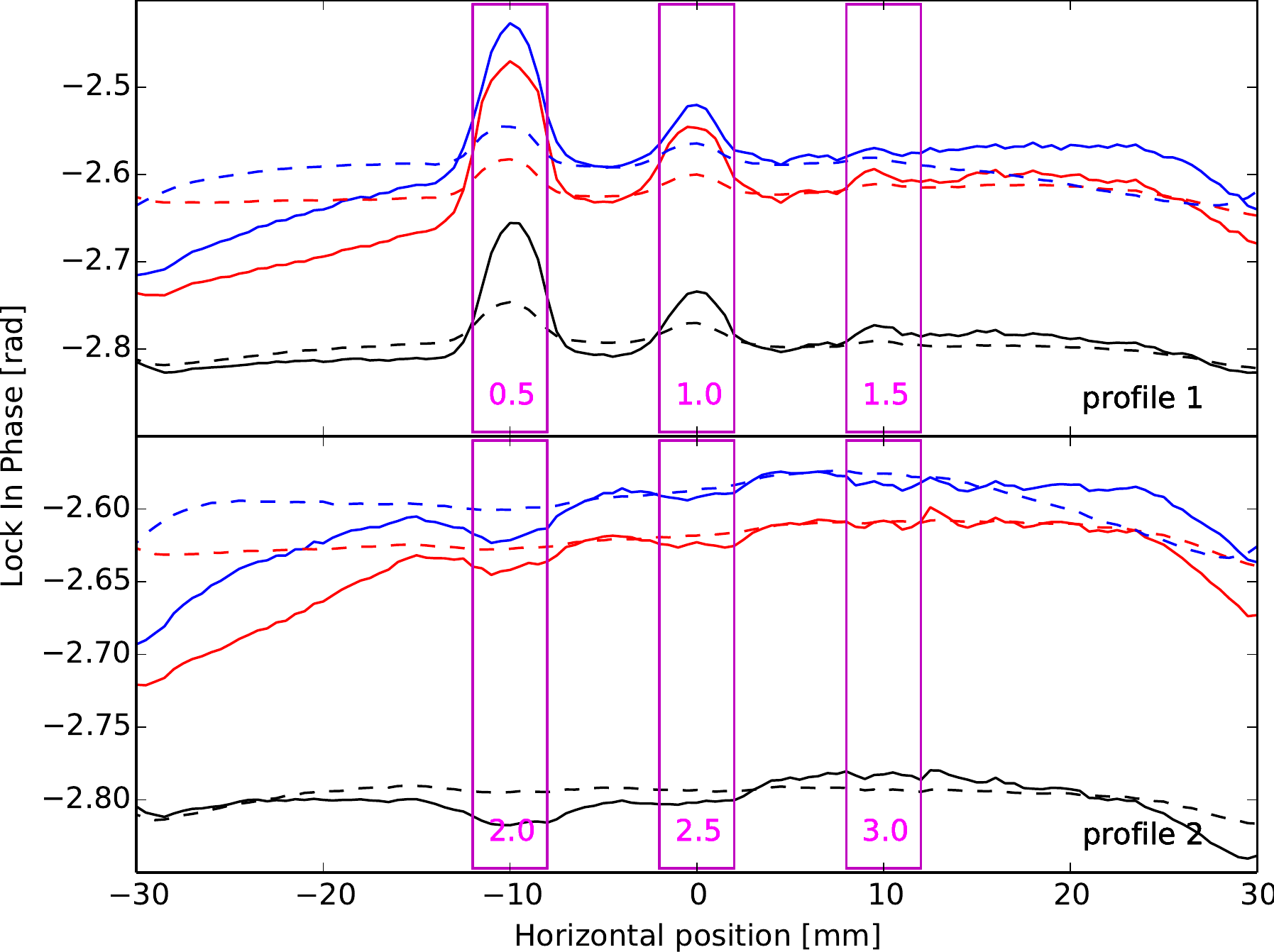}
\end{center}
\caption{Lock-in phase signals from measurements conducted at 0.01~Hz lock-in frequency, for 200~seconds, for ABS samples shown in Figure~\ref{fig:printedsample1}. \textcolor{black}{Black}, \textcolor{red}{red} and \textcolor{blue}{blue} lines represent data for low, medium and high power. Corresponding dashed lines were obtained from constant heating (not modulated) such that the mean heating would be the same, and processed using the same lock-in frequency. These lines were shifted upwards by 0.3 (\textcolor{black}{black}), 0.48 (\textcolor{red}{red}), and 0.50 (\textcolor{blue}{blue}) respectively to compare with results from modulated heating. Data is presented for profiles 1 and 2. \textcolor{magenta}{Magenta} boxes  represent the locations of known defects, with numbers indicating the depth of defects.
\label{fig:phasematpowjustify}}
\end{figure}

\begin{figure} 
\begin{center}
\includegraphics[width=\figwidth mm]{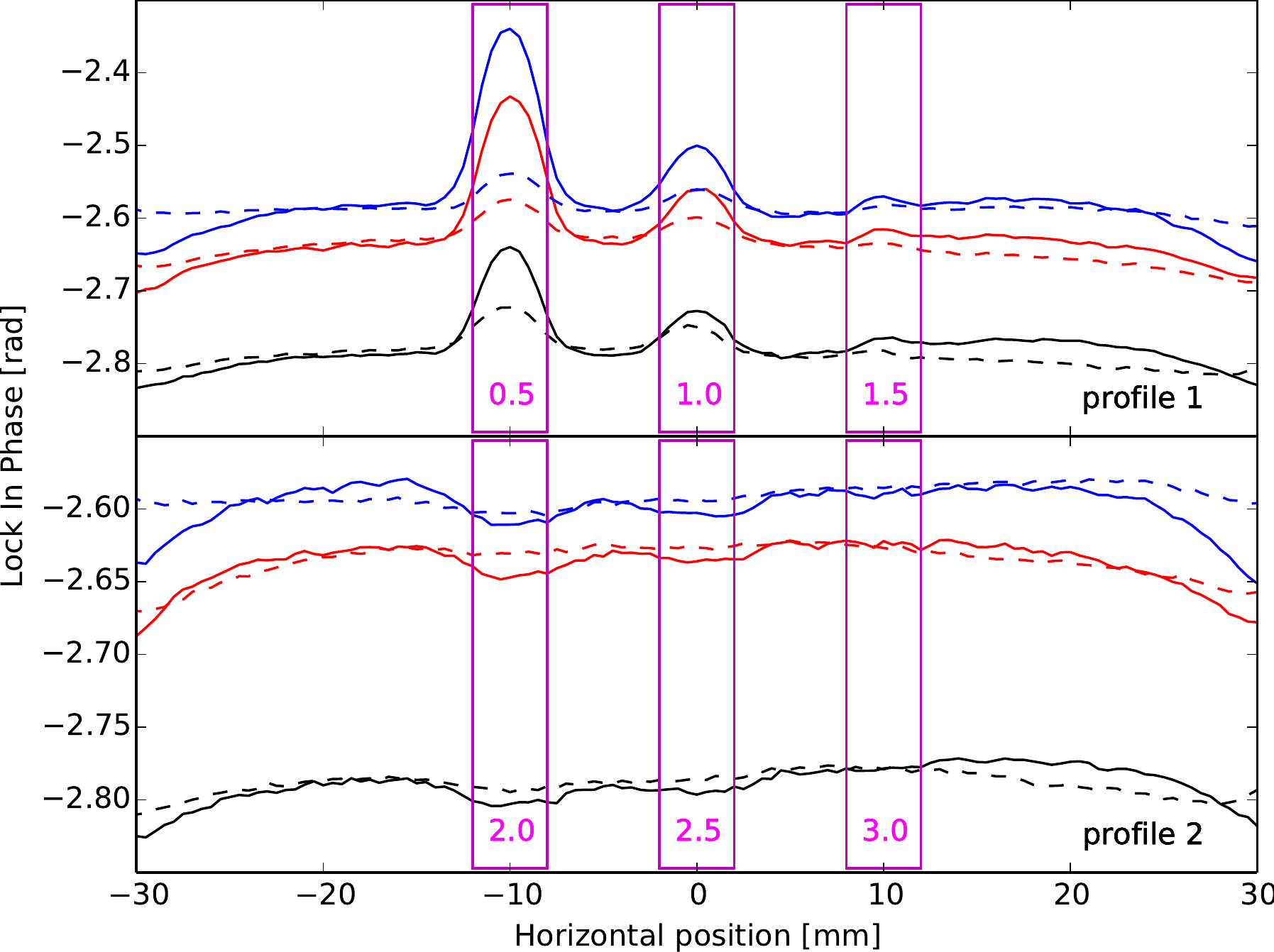}
\end{center}
\caption{Lock-in phase signals from measurements conducted at 0.01~Hz lock-in frequency, for 200~seconds, for ABS samples shown in Figure~\ref{fig:printedsample1}, using Ti55FT. \textcolor{black}{Black}, \textcolor{red}{red} and \textcolor{blue}{blue} lines represent data for low, medium and high power. Corresponding dashed lines were obtained from constant heating (not modulated) such that the mean heating would be the same, and processed using the same lock-in frequency. These lines were shifted upwards by 0.24 (\textcolor{black}{black}), 0.41 (\textcolor{red}{red}), and 0.42 (\textcolor{blue}{blue}) respectively to compare with results from modulated heating. Data is presented for profiles 1 and 2. \textcolor{magenta}{Magenta} boxes  represent the locations of known defects, with numbers indicating the depth of defects.
\label{fig:phasematpowTi55justify}}
\end{figure}

It is known that Lock-In thermography enhances the detection of defects by enhancing the signal to noise ratio, as proposed by Kuo \etal~\cite{Kuo1989SPIE}. In order to support that claim, benchmarking experiments were conducted, using non modulated heating with the same corresponding mean heating power, and compared with the results as shown before in Figures~\ref{fig:phasematpow} to \ref{fig:phasematpowoffsetTi55}. The results for ABS samples are shown in Figures~\ref{fig:phasematpowjustify} and~\ref{fig:phasematpowTi55justify} for SC7500 and Ti55FT respectively, where it is clearly shown that the signal contrast due to defects are amplified by modulated heating, with around 2 to 3 times the signal contrast if constant heating were applied to the samples.

\begin{figure} 
\begin{center}
\includegraphics[width=\figwidth mm]{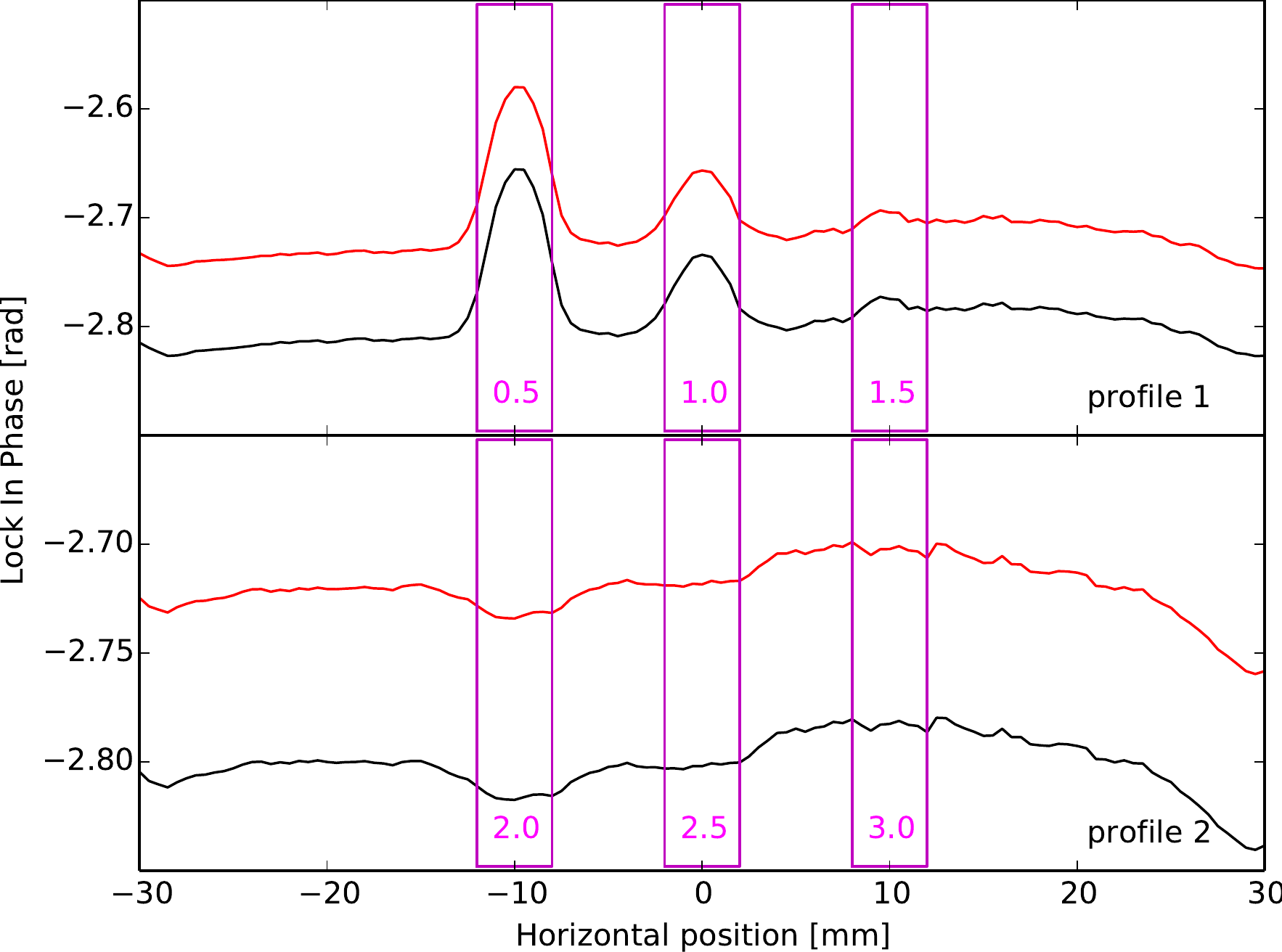}
\end{center}
\caption{Lock-in phase signals from measurements conducted at 0.01~Hz lock-in frequency, for 200~seconds, for ABS samples shown in Figure~\ref{fig:printedsample1}. \textcolor{black}{Black} and \textcolor{red}{red} lines represent data for synchronous (interpolated images) and asynchronous Lock-In Thermography. Data is presented for profiles 1 and 2. \textcolor{magenta}{Magenta} boxes  represent the locations of known defects, with numbers indicating the depth of defects.
\label{fig:phasematsync}}
\end{figure}

\begin{figure} 
\begin{center}
\includegraphics[width=\figwidth mm]{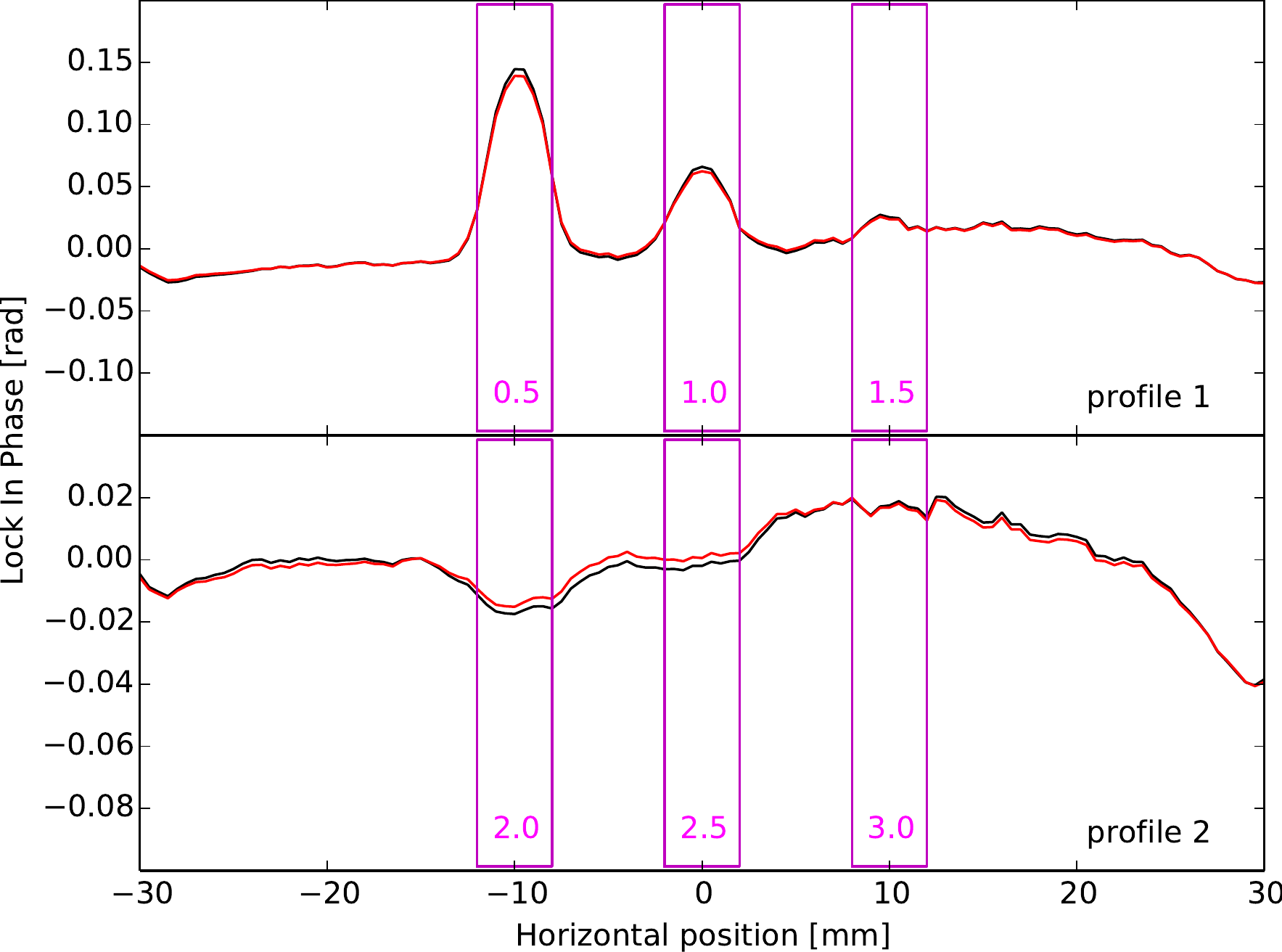}
\end{center}
\caption{Lock-in phase signals from measurements conducted at 0.01~Hz lock-in frequency, for 200~seconds, for ABS samples shown in Figure~\ref{fig:printedsample1}. \textcolor{black}{Black} and \textcolor{red}{red} lines represent data with baseline correction for synchronous (interpolated images) and asynchronous Lock-In Thermography. Data is presented for profiles 1 and 2. \textcolor{magenta}{Magenta} boxes  represent the locations of known defects, with numbers indicating the depth of defects.
\label{fig:phasematsyncoffset}}
\end{figure}

Given that the cameras did not have constant frame rates, it was interesting to investigate the relative effects of synchronised (interpolated) and non-synchronised images on the Lock-In phase results. These are shown in Figure~\ref{fig:phasematsync} for the ABS sample. As the Lock-In results were calculated using summation functions based on the signals, the values of $S_{0}$ and $S_{-90}$ would only be accurate if the frame rate was constant. This effect is shown by the obvious baseline shift when using synchronised images. However, the differences in the relative phase contrast were insignificant, as shown by the shifted curves in Figure~\ref{fig:phasematsyncoffset}. Nevertheless, it is important to generate Lock-In phase data from synchronised images, as the Lock-In equations are based on constant frame rate. Parameters $S_{0}$ and $S_{-90}$ are areas under a function (assuming constant time between frames), derived from Fourier Transform concepts, so deviations from constant frame rates would lead to deviations in Lock-In phase.

\begin{figure} 
\begin{center}
\includegraphics[width=\figwidth mm]{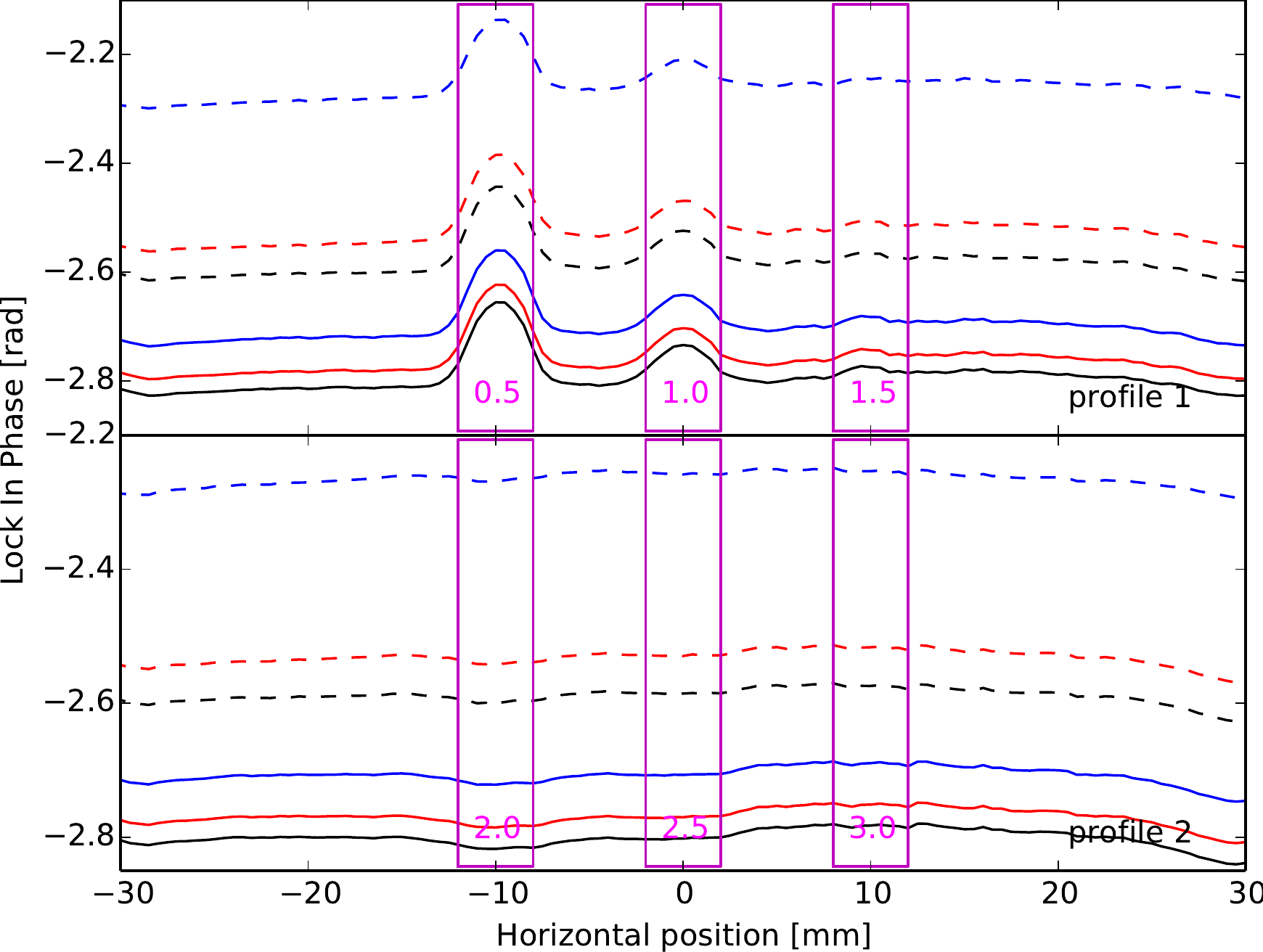}
\end{center}
\caption{Lock-in phase signals from measurements conducted at 0.01~Hz lock-in frequency, for 200~seconds, for ABS samples shown in Figure~\ref{fig:printedsample1}. Full sampling (\textcolor{black}{Black}) was compared with subsampling of every 2 (\textcolor{red}{red}), 4 (\textcolor{blue}{blue}), 8 (\textcolor{black}{black dashed}), 10 (\textcolor{red}{red dashed}) and 20 (\textcolor{blue}{blue dashed}) frames from the full data. Data is presented for profiles 1 and 2. \textcolor{magenta}{Magenta} boxes  represent the locations of known defects, with numbers indicating the depth of defects.
\label{fig:phasematsubsamp}}
\end{figure}

\begin{figure} 
\begin{center}
\includegraphics[width=\figwidth mm]{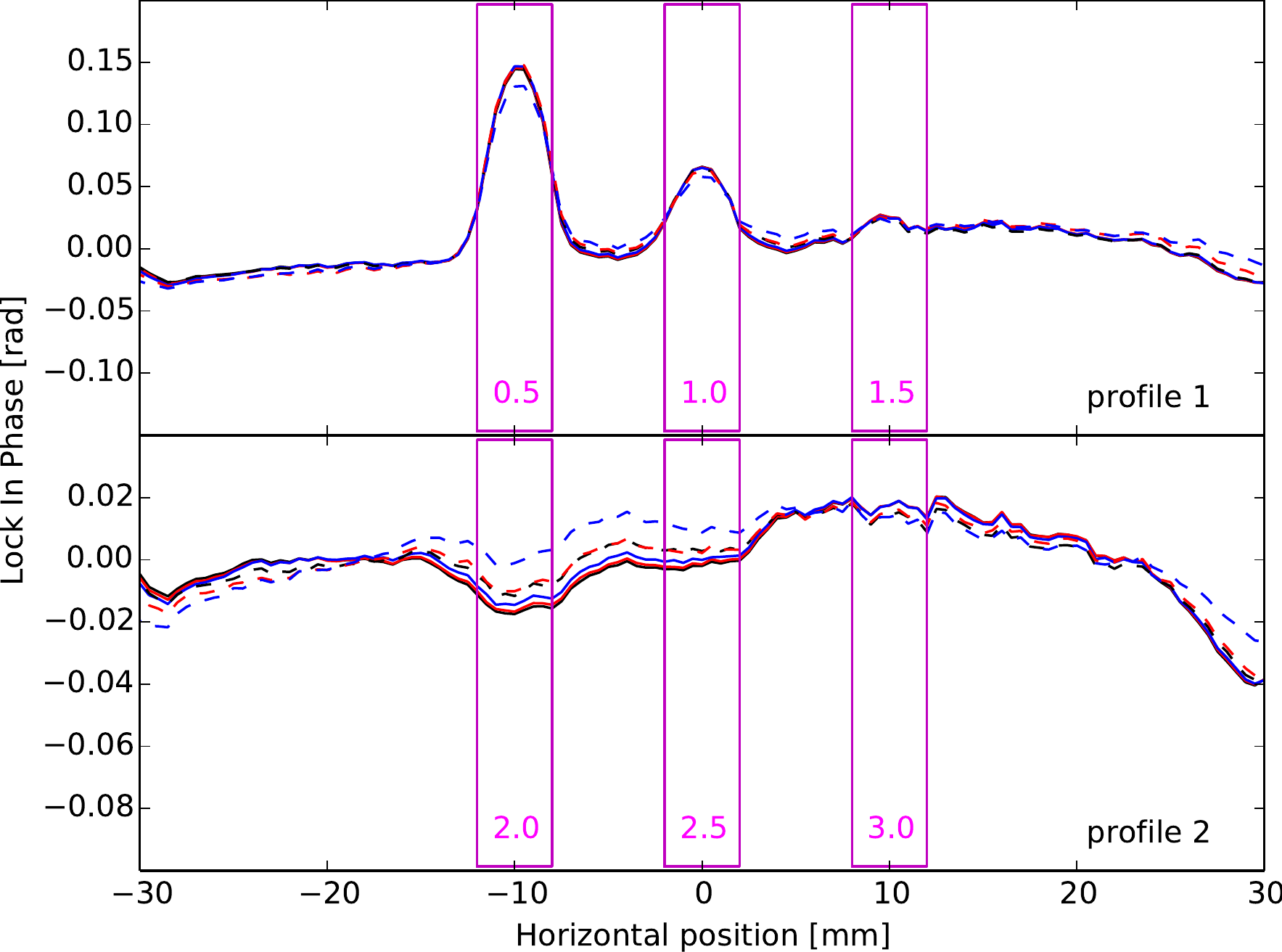}
\end{center}
\caption{Lock-in phase signals from measurements conducted at 0.01~Hz lock-in frequency, for 200~seconds, for ABS samples shown in Figure~\ref{fig:printedsample1}. Full sampling (\textcolor{black}{Black}) was compared with subsampling of every 2 (\textcolor{red}{red}), 4 (\textcolor{blue}{blue}), 8 (\textcolor{black}{black dashed}), 10 (\textcolor{red}{red dashed}) and 20 (\textcolor{blue}{blue dashed}) frames from the full data, with baseline correction of all data. Data is presented for profiles 1 and 2. \textcolor{magenta}{Magenta} boxes  represent the locations of known defects, with numbers indicating the depth of defects.
\label{fig:phasematsubsampoffset}}
\end{figure}

In order to further understand the relative effects of the frame rate on Lock-In phase, sub-sampling was implemented on a single measurement using ABS, with various sub-sampling parameters (2,4,8,10,20). For instance, every 10 images were taken to generate the Lock-In results if sub-sampling parameter was set to 10. The corresponding Lock-in phase data is shown in Figure~\ref{fig:phasematsubsamp}, where it is clearly shown that the camera frame rate would affect the baseline result of the Lock-In phase. Baseline offset was applied to these results, and shown in Figure~\ref{fig:phasematsubsampoffset}. The shifted profiles show that significant deviations would be observed if the measurements were done at frame rates lower than 0.1~$s^{-1}$, corresponding to sampling every 10 frames, or 10 frames per Lock-In period, which is more than twice the minimum recommended value of 4 by Bauer \etal~\cite{Bauer2009EDFA}. These profiles also show that the signal contrast is compromised with frame rates lower than 0.1~$s^{-1}$, for defects at or deeper than 2.0~mm below the surface.

\begin{figure} 
\begin{center}
\includegraphics[width=\figwidth mm]{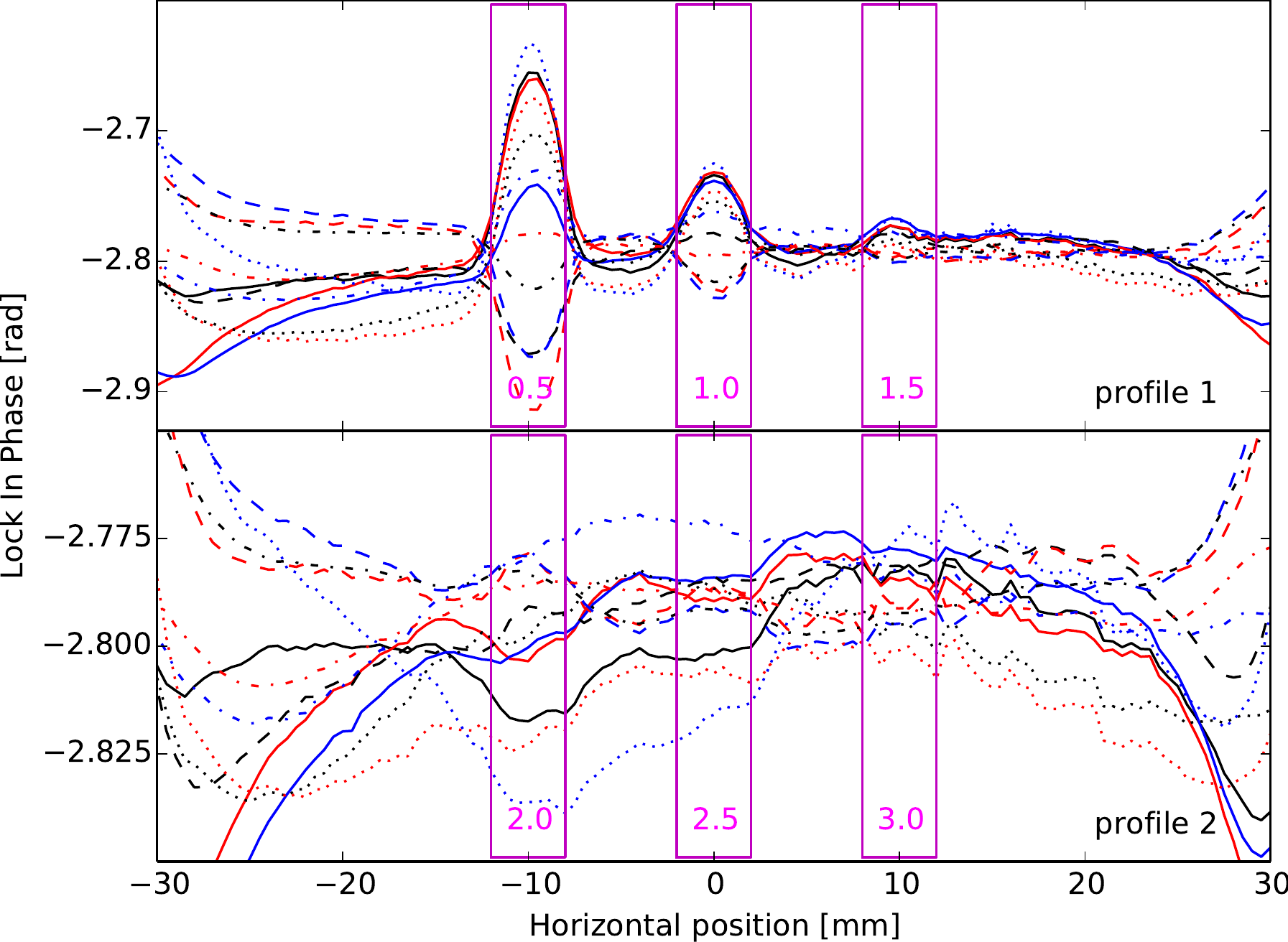}
\end{center}
\caption{Lock-in phase signals from measurements conducted at 0.01~Hz lock-in frequency, for 200~seconds, for ABS samples shown in Figure~\ref{fig:printedsample1}. Full sampling (\textcolor{black}{Black}) was compared with subsampling of 100 images with first frame equal to 0~(\textcolor{red}{red}), 10~(\textcolor{blue}{blue}), 20~(\textcolor{black}{black dashed}), 30~(\textcolor{red}{red dashed}), 40~(\textcolor{blue}{blue dashed}), 50~(\textcolor{black}{black dashed dotted}), 60~(\textcolor{red}{red dashed dotted}), 70~(\textcolor{blue}{blue dashed dotted}), 80~(\textcolor{black}{black dotted}), 90~(\textcolor{red}{red dotted}), 100~(\textcolor{blue}{blue dotted}) frames from the first frame, with baseline correction of all data. Data is presented for profiles 1 and 2. \textcolor{magenta}{Magenta} boxes  represent the locations of known defects, with numbers indicating the depth of defects.
\label{fig:phasematsubsampstarttime}}
\end{figure}

From a single measurement with 200 frames, it was possible to investigate the possible minimum number of frames required to get reasonable phase contrast. The relative start time of the measurement was also of interest. Hence, 100 sequential frames were processed to compare with phase data for the full measurement of 200 frames. The first frame was also varied in steps of 10 to investigate the feasibility of measuring halfway through the heating process. Note that the Lock-In frequency was set to 0.01~Hz for the ABS sample, and the full measurement covered 2 complete cycles. The corresponding baseline corrected Lock-In phase results are shown in Figure~\ref{fig:phasematsubsampstarttime}. These profiles show that for the shallower defects up to 1.5~mm in depth, the first 100 frames and the last 100 frames were sufficient, as they showed similar signal contrast. However, for defects from 2.0~mm depth, the phase contrast was more prominent for the full measurement. Hence, even though it is possible to reduce the measurement time, it may not enable the reliable detection of deeper defects. Also, as the first frame varied in steps of 10, it could be observed that the relative phase changed significantly. These findings show that for a measurement to start halfway through heating, the first frame has to be in sync with the heating cycle for consistent results. In other words, if measurements were to start halfway through the heating process, the first frame has to be acquired in multiples of 100~s for Lock-In frequency of 0.01~Hz, or multiples of 50~s for 0.02~Hz. 

\section{Conclusion}
\label{sect:conclude}

The current work explored many variables to understand the relative effects on Lock-In thermography, on 3D printed components. These include print direction of samples, materials (ABS and PLA), cameras, heating power, lock in frequency, as well as thickness, width and depth of defects. Lock-In thermography was also benchmarked against constant heating, and showed up to around 2-3 times amplification of phase contrast, which allowed defect locations to be more visible. It was shown that print direction resulted in baseline shift in Lock-In phase signals, but relative phase contrast was not affected. Between ABS and PLA, there was similar contrast, but different baselines in the phase signal. Different cameras had different baseline signals, with similar contrast, and it was shown that cheap cameras such as Seek Thermal CompactXR were capable of detecting the same defects that were detected on other cameras. Phase contrast did not get affected by higher heating power, except for shallower defects up to 1.0~mm depth, where higher contrast was observed with higher heating power. As expected, deeper defects could be detected with lower values of Lock-In frequency. In addition, wider, thicker and shallower defects resulted in better phase contrast. The phase signals for defects that were 4~mm wide showed that it was possible for automatic detection of defects up to around 1.5~mm in depth, as phase signals were similar for deeper defects at different depths. Sub-sampling was also conducted on a single set of data, and it was found that the minimal number of frames per Lock-In period was 10 for minimal deviations in the phase contrast. Also, phase contrast results were similar for defects up to 1.5~mm in depth with data from just 1 Lock-In period, compared to the full measurement of 2 Lock-In periods, as long as the first frame was synchronised with the heating cycle.


\section*{References}
\label{sect:references}

\bibliography{thermographybib}

\begin{table}[h]
\caption{Specifications of cameras used, as well as experimental conditions.
\label{tab:cameras}}
\begin{indented}
\item[]\begin{tabular}{@{}*{5}{l}}
\br
Camera & Waveband & NETD & Frame & Trigger\\
 & & & Rate & \\
~[-] & [$\mu m$] & [$mK$] & [$s^{-1}$] & [-]\\
\mr
FLIR SC7500 & 1.5-5.1 & $<$25 & 1 & TTL\\
FLIR A310 & 7.5-13 & 50 & 2 & Time\\
Fluke Ti55FT & 8-14 & 50 & 0.5 & N.A.\\
Seek Thermal & 8-14 & Unknown & 1 & N.A.\\
CompactXR & & & & \\
Xenics Gobi 640 & 8-14 & 55 & 1 & N.A.\\
\br
\end{tabular}
\end{indented}
\end{table}

\end{document}